\documentclass[12pt]{iopjournal}
\usepackage[numbers]{natbib}
\usepackage{comment}
\usepackage{ragged2e}
\usepackage{amsmath}
\usepackage{float}
\usepackage{lmodern}

\begin{document}


\title{Structure and dynamics of interface around a quenched impurity in a
colloidal liquid film}

\author{Sabuj Mandal$^{1,*}$\orcid{0009-0008-9068-179X} and Jaydeb Chakrabarti$^{1}$\orcid{0000-0003-0753-3259}}

\affil{$^1$Department of Physics of Complex Systems, S. N. Bose National Centre for Basic Sciences, Salt Lake, JD Block, Sector 3, Kolkata-700106, India}


\affil{$^*$Author to whom any correspondence should be addressed.}

\email{{sabuj.mandal605@gmail.com}}

\justifying
\begin{abstract}
\justifying
The interfacial structure and dynamics around a quenched impurity is far from understood till date. Here we explore the structural and dynamical properties of interface between colloidal fluid film and  a quenched impurity by using computer simulation. We tune the size($\sigma_{imp}$) of a quenched pinned impurity and observe that the impurity surface is wetted by a fluid layer, giving rise to a finite-width interfacial region for sufficiently large impurities. Beyond this interfacial region, the host fluid exhibits quasi-long ranged orientational order(QLRO) due to the fluid- crystal phase coexistence. We find that the mean diffusivity becomes very low in the QLRO phase where the self van Hove function shows an exponential tail, signature of dynamic heterogeneity within the system. In the presence of second impurity the orientational order is suppressed if the fluid layers around the impurity particles overlap.
\end{abstract}

\section{Introduction}

Impurities drastically affect condensed phase properties. The impurity in a host matrix creates interface where foreign atoms or contaminants meet the host phase. This interface dictates how defects localized at the interface interact with the bulk material, governing performance to control industrial processes\cite{r31,r32,r34,r33, guruswamy2022}, ranging from nano-electronics\cite{r30} to bio-pharmaceuticals\cite{r29}. Impurities are pedagogically interesting as well\cite{r42}:
On one hand, impurities are known to destroy the long ranged order in an ordered system and replace it by a quasi-long ranged order\cite{rjc,r65}. On the other hand, impurities help a system to overcome nucleation barrier for crystallization from a fluid phase\cite{Frenkel2005,r76}. Despite both technological importance and pedagogical interests, host-impurity interface is  poorly understood to date to the best of our knowledge.

Colloid dispersions are ideal systems to study impurity induced changes in condensed phase properties at microscopic level by following the particle motion in the laboratory\cite{r5,r6}. Experimental studies \cite{tsiok2025springer,sun2021cpl} on two-dimensional colloidal suspensions with randomly pinned particles demonstrate that quenched disorder can fundamentally alter the equilibrium phase behavior replacing the ordered crystals by hexatic and glassy phases as the pinning fraction increases. Experiments on two-dimensional colloidal crystals containing anisotropic impurities \cite{chen2021prl, gray2015jpcm,takae2013aps,zheng2011prl} reveal that  localized defects are generated around isolated impurities at low concentrations, whereas at higher concentrations these defects percolate to form grain boundaries which break the crystal into polycrystalline domains. 

Crystallization from a fluid phase has been also extensively investigated in the presence of flat substrates~\cite{r16,r17,r18}, particle assemblies~\cite{r44,r45}, structured surfaces~\cite{r46,r47}, and external seed particles or impurities~\cite{Lowen2015,r12,Frenkel2005,r15} in a colloidal fluid. Recent MD simulations \cite{zhao2024elsevier} report the effects of nano-meter sized impurities on liquid-solid transitions in water droplets. Impurities scattered in the droplet creates void inside the frozen droplet, while clusters of impurity particles create a uniform ice structure. Another Monte Carlo study\cite{ramesh2025jcp} reports on the effects of adding  impurities in a metastable phase with coexisting fluid-FCC phases. It is observed that impurities with BCC, SC, or amorphous rough surface induce interfacial BCC structure and  drives the bulk system into a stable BCC phase. FCC impurity seeds, on the other hand, creates FCC like structures in the interface but in the equilibrium  a BCC phase appears within the system. This study highlights that the host-impurity interface behaving distinctly from the stable bulk phase  depending on the impurity structure. 

With this backdrop we carry out molecular dynamics (MD) simulations to study quenched impurity-host interfacial structure and dynamics in a colloidal fluid in two dimensions. Unlike  the previous study\cite{ramesh2025jcp}, 
we take the host system in a fluid phase and dope it with a quenched impurity.
Previous studies show\cite{jc1995prl} that the fluid-crystal phase transition changes behavior qualitatively if the colloidal fluid is modulated by external modulation near the bulk transition point. This makes us to choose the bulk fluid phase near the bulk fluid-crystal phase transition. The colloidal particles interact via the repulsive screened Coulomb interaction potential with a given inverse Debye screening length $\kappa$\cite{r43}. We consider quenched impurity by immobilizing a randomly selected fluid particle. The impurity particle interacts with the other particles via the screened coulomb repulsive potential with the same inverse screening length as the other colloidal particles. We tune the size of the pinned particle while keeping the packing fraction and the screening length of the host-impurity interaction the same.

We calculate from the equilibrium configurations the colloidal fluid density profile and bond order parameter profile around the impurity particle corresponding to 6-fold coordination, $Q_6$ depicting the crystalline order, 5-fold coordination, $Q_5$ and 7-fold coordination $Q_7$ describing the defect sites, the 6-fold bond orientation correlation function $g_6(r)$ and the 6-fold order parameter susceptibility $\chi_{6}$. We observe that the impurity particle is covered by a fluid layer of high local number density but having low $Q_6$.   The bond order parameters saturate beyond an interfacial width. $Q_6$ both within the interface and the saturated region and the interfacial width grow continuously with impurity size. The interface has high number of defects with 5- and 7-fold coordination which proliferate into region far from the impurity surface. This results in quasi-long ranged order (QLRO) with slow algebraic decay of $G_6(r)$. QLRO is a fluid-crystal phase coexistence as revealed by the $\chi_{6}$ peaks.  The particle dynamics show heterogeneity: particularly, the interfacial particles move faster over the impurity surface.

\section{ Methods: Model and simulation details}

The interaction in the system are described the repulsive part of the DLVO potential, also known as electric double layer potential \cite{safran1994book} :
\begin{equation}
 V(r)= 
      \begin{cases}
     \infty, \;\;\;\;\; r<\sigma\\
     \frac{V_{0}}{\kappa}exp[-\kappa (r -\sigma_{ij})], \;\;\;\;\; \sigma<r<r_{c} \\
      0, \;\;\;\;\;\; r>r_{c} \\
      \end{cases}
\end{equation}\\
We study the system using MD simulations in a box under periodic boundary conditions in two dimensions as implemented in the Large-scale Atomic/Molecular Massively Parallel Simulator (LAMMPS) \cite{r19} software at constant NVT. Here $\sigma$ is taken as the length unit. The energy unit is $k_BT$ at room temperature ($T$=300K) and $m (= 10^{-16} kg)$, the mass of a colloidal particle, the mass unit. The system temperature $T (=k_{B}T/V_{0})$ is maintained by the Nosé-Hoover thermostat at a constant of value 1.0 throughout the simulations.

We study the system of N (= 9408) colloidal particles within an area ($A$) of rectangular surface ($L_{x}=90 \sigma$ and $L_{y}=103.9\sigma$). The box dimensions are chosen as $L_{y}= \frac{\sqrt{3}}{2} L_{x}$ so that the triangular lattice formation is not frustrated. This corresponds to the packing fraction, $\phi(=\frac{N\pi }{4L_xL_y})$=0.79. The integration are performed using a time step of 0.001. The total run length of the simulations is  $2 \times 10^6$ MD steps. The equilibration of the system is monitored by the potential energy of the system. All the data are taken after the system gets equilibrated for $5 \times10^5$ MD time steps.

First, we study the system without any impurity. After equilibration of the bulk we randomly pin a particle, called as quenched impurity and change its size systematically, keeping the density of the bath particles same as in bulk. To avoid the particles overlapping, we perform energy minimization after introducing the immobile particle prior to the MD runs. We repeat the simulations for four different positions of the quenched particle. 

We first calculate radial distribution function to study the structure of the system given by
\begin{equation}
 g(r) =\langle\frac{1}{N} \sum_{ij} \delta(r -|r_i -r_j|)\rangle
\end{equation}
where $r_i$ is the position of first particle and second particle located at a distance $r_j$. $\langle...\rangle$  is the average over configurations.
We calculate the bond orientation order parameter of the of the order $l$ for $j$-th particle\cite{r21,r66}: 
\begin{equation}
\psi_{l}(\mathbf{r}_j)=\langle\frac{1}{N_j} \sum_{k=1}^{N_j} exp(il\theta(\mathbf{r}_{jk}))\rangle
\end{equation}
where $\langle...\rangle$ is the average over the configurations and the trajectories. $l$ is an integer, varying from 1 to 6. Here $l$=6 is relevant for a hexagonal 2D lattice system. The summation $k$ runs over $N_j$ neighboring particles of a given particle within a cutoff distance that is taken up to the first minimum of the radial distribution function. The angle between the bond vectors connecting the given particle with its neighboring particle and a reference axis(say x) is defined as $\theta(\mathbf{r}_{jk})$, where $\mathbf{r}_{jk}$ is the line joining between the particles j and k.


We calculate the specific heat ($C_{V}$) of the system from the energy fluctuations\cite{r40}:
\begin{equation}
C_{V}=\frac{\langle(E-\overline{E})^2\rangle}{k_{B}T^2} 
\end{equation}\\
where $E$ and $\overline{E}$ are the total and mean energies of the system. 

We further examine the bond orientation correlation function\cite{r20} $G_6(\mathbf{r})$ defined  as
\begin{equation}
G_6(r)=\langle(\psi_{6}^*(\mathbf{r}_i)\psi_{6}(\mathbf{r}_j))\rangle
\end{equation}
where $\mathbf{r}= \mathbf{r}_i -\mathbf{r}_j$, the distance between two points  $\mathbf{r}_i$ and  $\mathbf{r}_j$ in the system.

We also calculate the bond order susceptibility\cite{r36,r67} defined as: 
\begin{equation}
\chi_6=\langle(Q_6)^2\rangle - \langle Q_6\rangle^2
\end{equation}
where $\langle...\rangle$ represents the ensemble average for each sub-blocks. The bond order susceptibility is calculated for various length scales by dividing the system into equal sub-blocks of length $L_b= L/4, L/8, L/16$ and $L/32$. For each sub-block we compute the total bond order parameter($Q_6$) for each configurations. We calculate $\chi_6$ from the fluctuations of $Q_6$ for each sub-blocks. Then the distribution of $\chi_6$,  $P(\chi_6)$ over the system is computed.

To study the dynamics of the system we compute the self part of the van Hove function\cite{r75}, defined as the distribution of displacement($\vec r)$ of particles in the time interval $\Delta t$, 
\begin{equation}
G_s(r,\Delta t)=\langle\frac{1}{N} \sum_i \delta[r -(\vec r_i(\Delta t) - \vec r_i(0))]\rangle
\end{equation}
where $\vec r_i(0)$ is the ith particle position at time, t=0 and $\vec r_i(\Delta t)$ is its position at a later time $\Delta t$. The average is over the configurations.

\section{Results and discussions}


First we study the 2D bulk colloidal system. We fix the density of colloidal particles  and vary the inverse screening length $\kappa$ in Eq.1.  We show in Figure 1(a), g(r) (Eq.2, Methods) which decays to unity after a few peaks for lower $\kappa(=5.5)$ value. So the bulk system is short range in translational order typical of a fluid phase. The bond order parameter($Q_6$), given by  average $\psi_{6}(\mathbf{r}_j)$ (Eq.3, Methods) over all the particles and  the specific heat($C_{v}$) (Eq.4, Methods) are shown in Figure 1(b) and 1(c) respectively. $Q_{6}$ in Figure 1(b) shows change from a value close to $Q_f$=0.58 (fluid) around $\kappa$=5.9 to a value exceeding $Q_T$ =0.8 (triangular lattice) around $\kappa$=7.0. We observe in Figure 1(c) that there is a sharp peak in $C_{V}$ at $\kappa$=5.9. We identify $\kappa$=5.9 as a fluid-hexagonal crystal transition point in 2D bulk system. We also show that $G_6$ (Eq. 5, Methods) in the bulk fluid phase ($\kappa=$5.5), slightly below the transition point, in Figure 1(d) we can see that the falls exponentially with r, characteristic of a fluid phase. Thus, the bulk system at this state point is a fluid with short ranged orientational and translational order.

\begin{figure}[H]
\centering
    \includegraphics[width=0.32\linewidth]{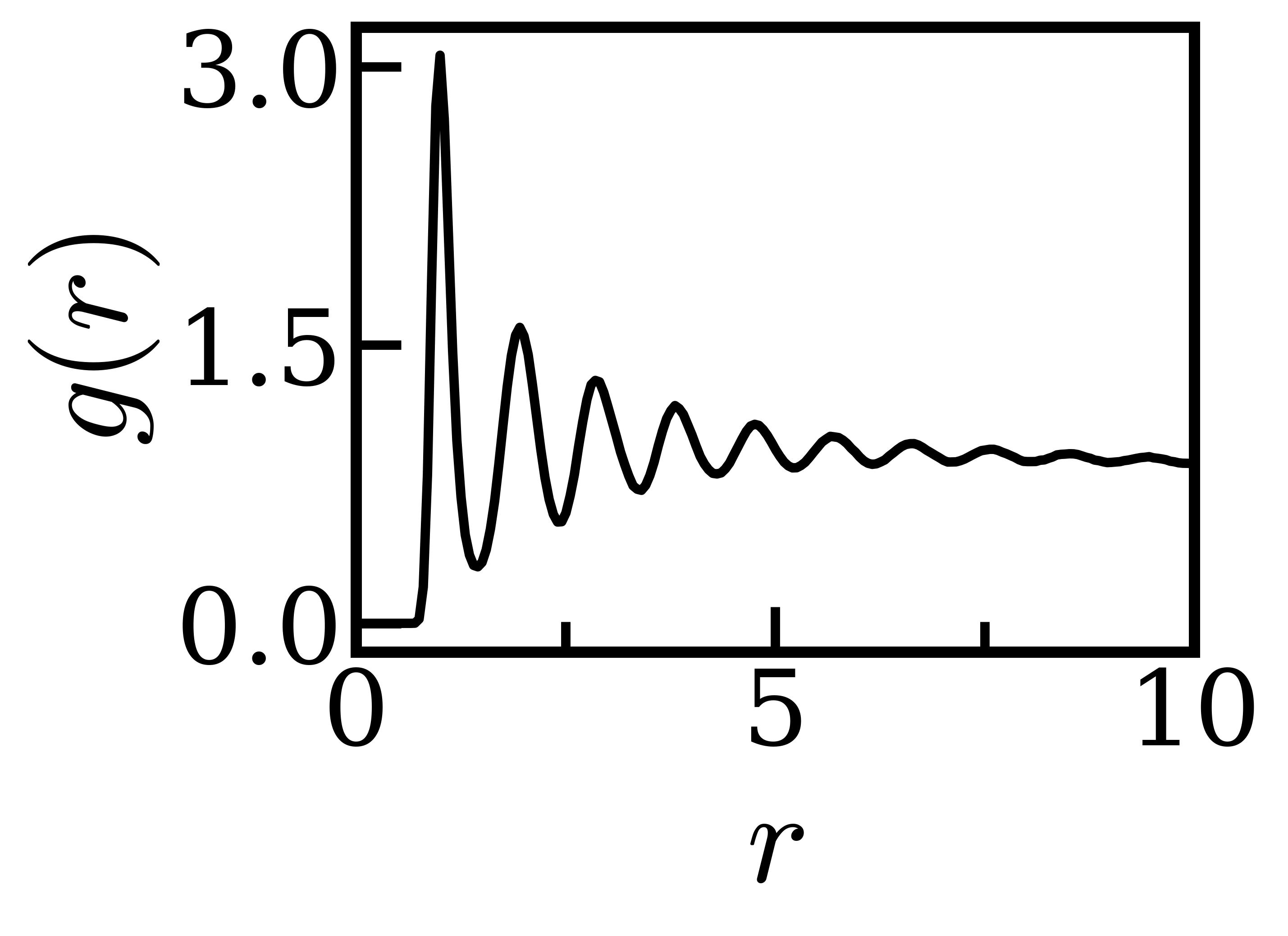}
        \put(-30,72){\textbf{(a)}}
    \hspace{0.2cm}\includegraphics[width=0.31\linewidth]{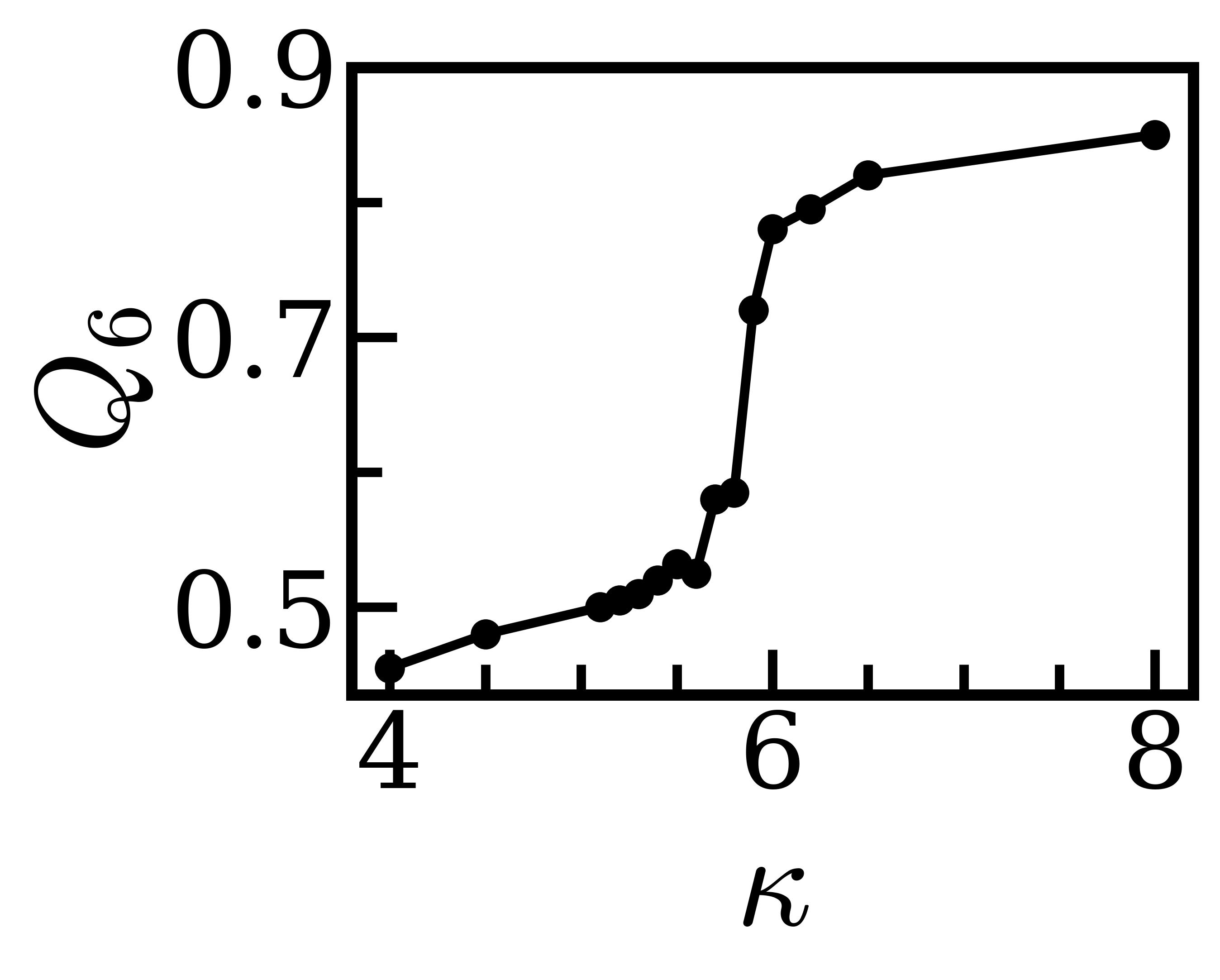}
        \put(-24,72){\textbf{(b)}}\\
    \includegraphics[width=0.32\linewidth]{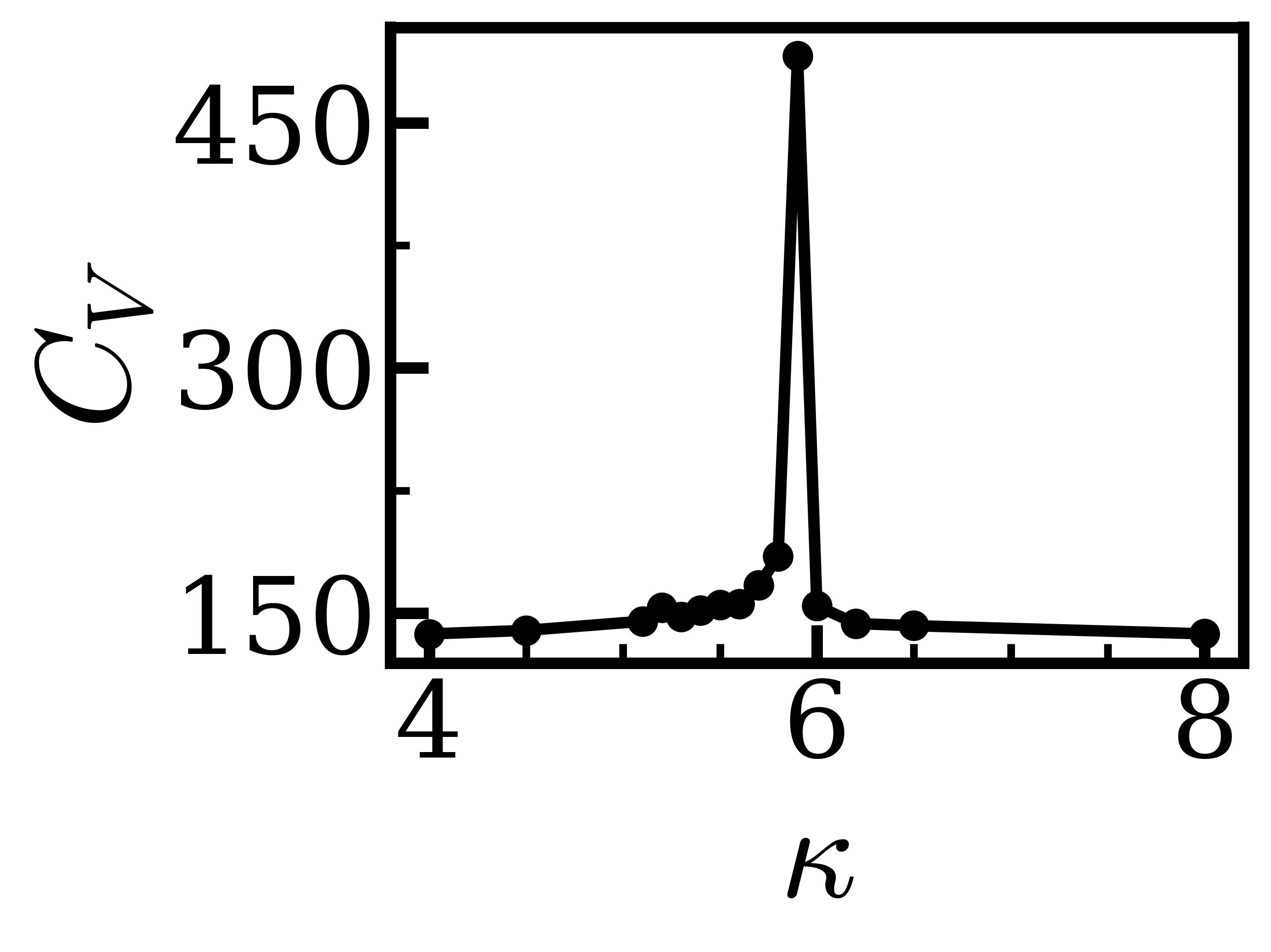}
        \put(-24,72){\textbf{(c)}} 
    \includegraphics[width=0.33\linewidth]{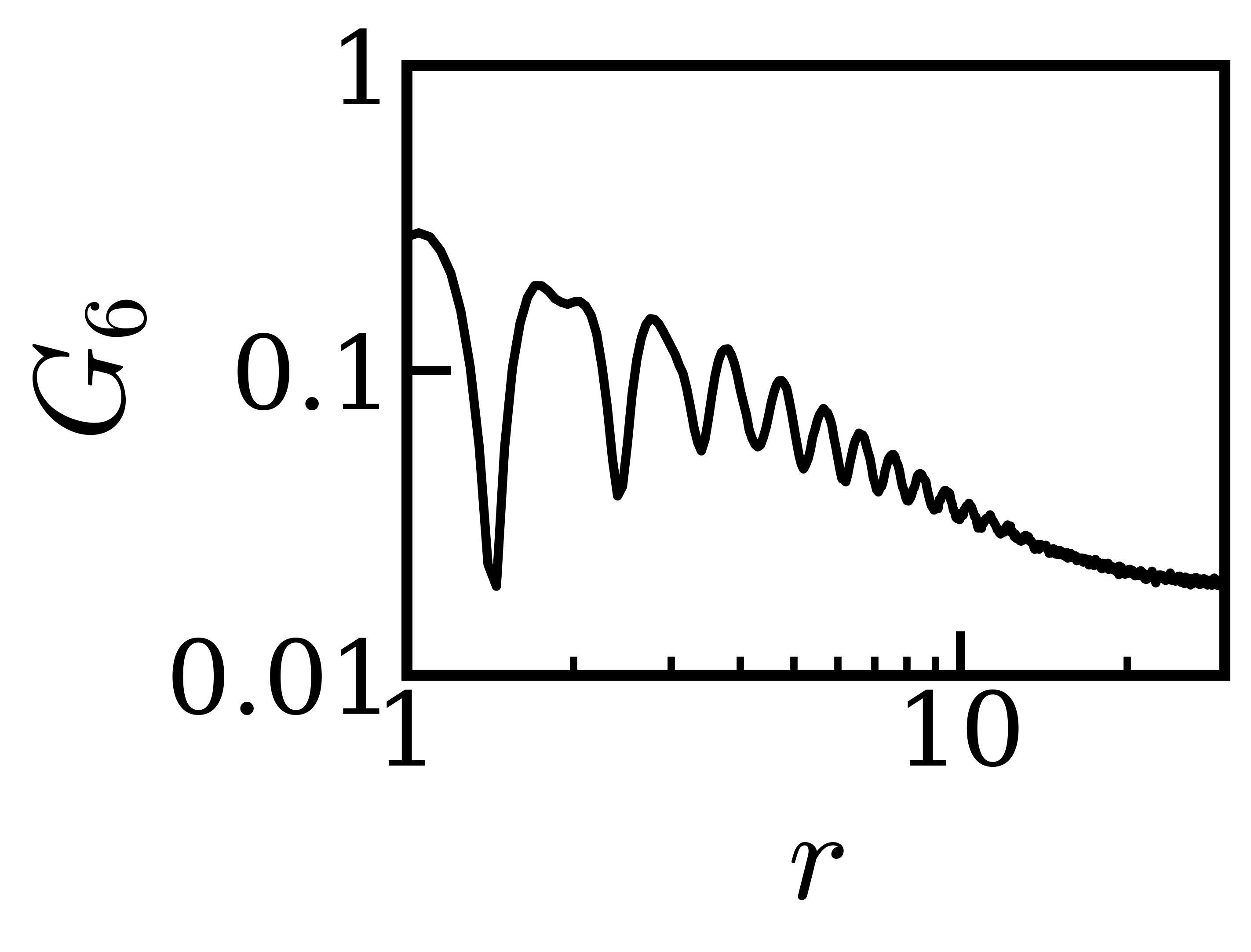}
        \put(-24,72){\textbf{(d)}}
    \caption{\small{Bulk data :  \textbf{(a)} Radial probability distribution function at $\kappa=5.5$. \textbf{(b)} Total bond order parameter ($Q_{6}$) vs inverse screening length($\kappa$). \textbf{(c)} Specific heat ($C_{v}$) vs  $\kappa$ plot. \textbf{(d)} $G_6$ vs r plot at $\kappa=5.5$.}}

\end{figure} 

\subsection{Interfacial structure}

Now we consider colloidal fluid at $\kappa(=5.5) $, below the bulk fluid-crystal transition.  The order parameter value at this 2D bulk phase point, $Q_{6}=$ 0.53 which corresponds to a fluid phase. We also calculate the $Q_5$ and $Q_7$, averages of $\psi_{l}(\mathbf{r}_j)$ corresponding to $l$=5 and 7 respectively. In this state, $Q_5$ and $Q_7$ are 0.31 and 0.41 respectively (data not shown). We immobilize a randomly selected particle such that the total number of particles (colloidal+impurity) remains the same. The impurity particle has the same interaction as with the bulk particles. We tune the diameter,  $\sigma_{imp}$ of the impurity particle. 

We show the density profile of the bath particles around the pinned impurity in Figure 2(a) for  $\sigma_{imp} =32$. The density profile exhibits pronounced oscillations near the impurity surface followed by a gradual decay toward the bulk density, indicating the high density around the impurity. The first density peak appears immediately outside the excluded-volume region and reaches a value significantly higher than the bulk density, reflecting particle accumulation near the impurity surface. 
\begin{figure}[H]
\centering
    \includegraphics[width=0.29\linewidth]{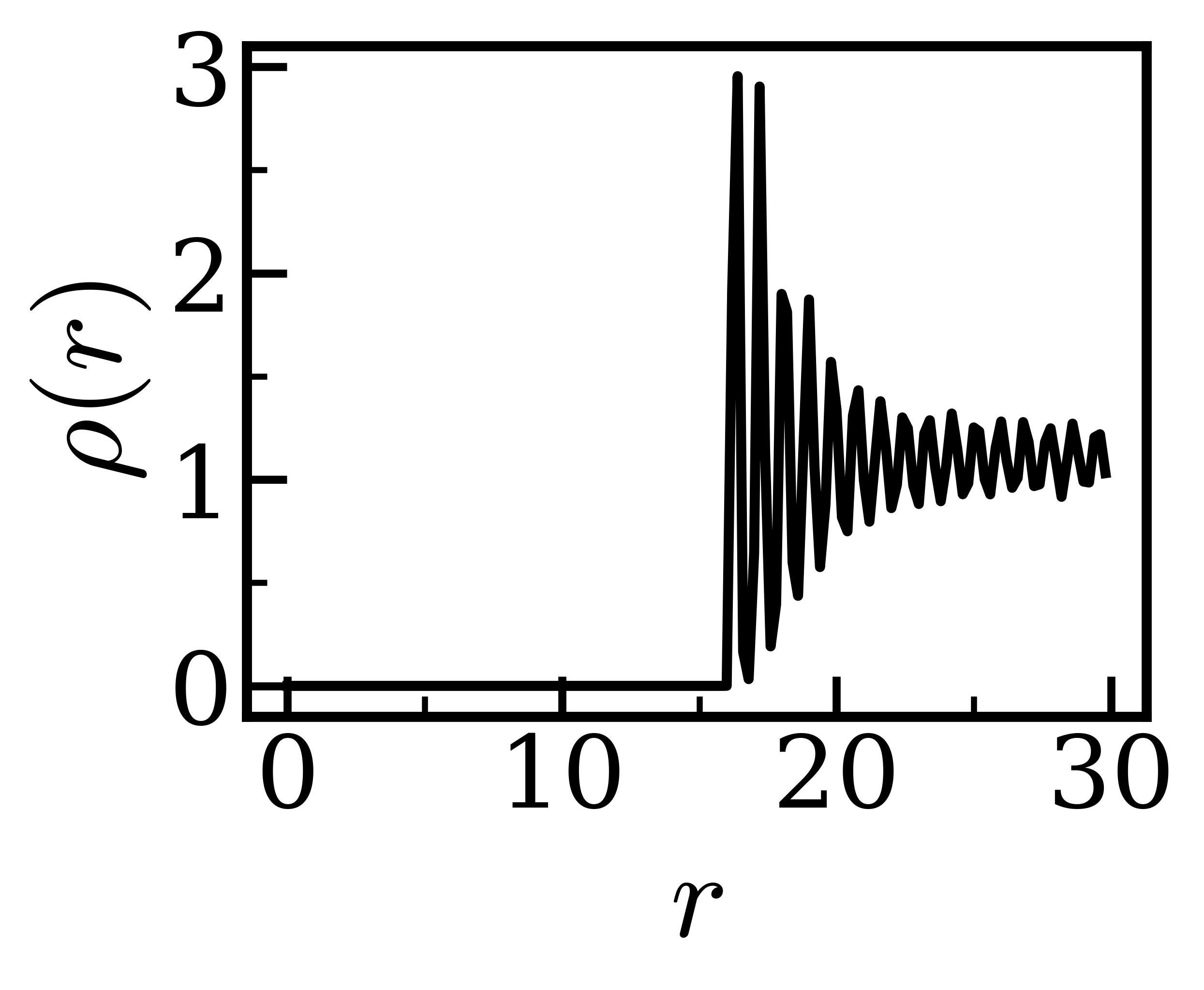}
        \put(-95,85){\textbf{(a)}} 
    \hspace{0.5cm}\includegraphics[width=0.31\linewidth]{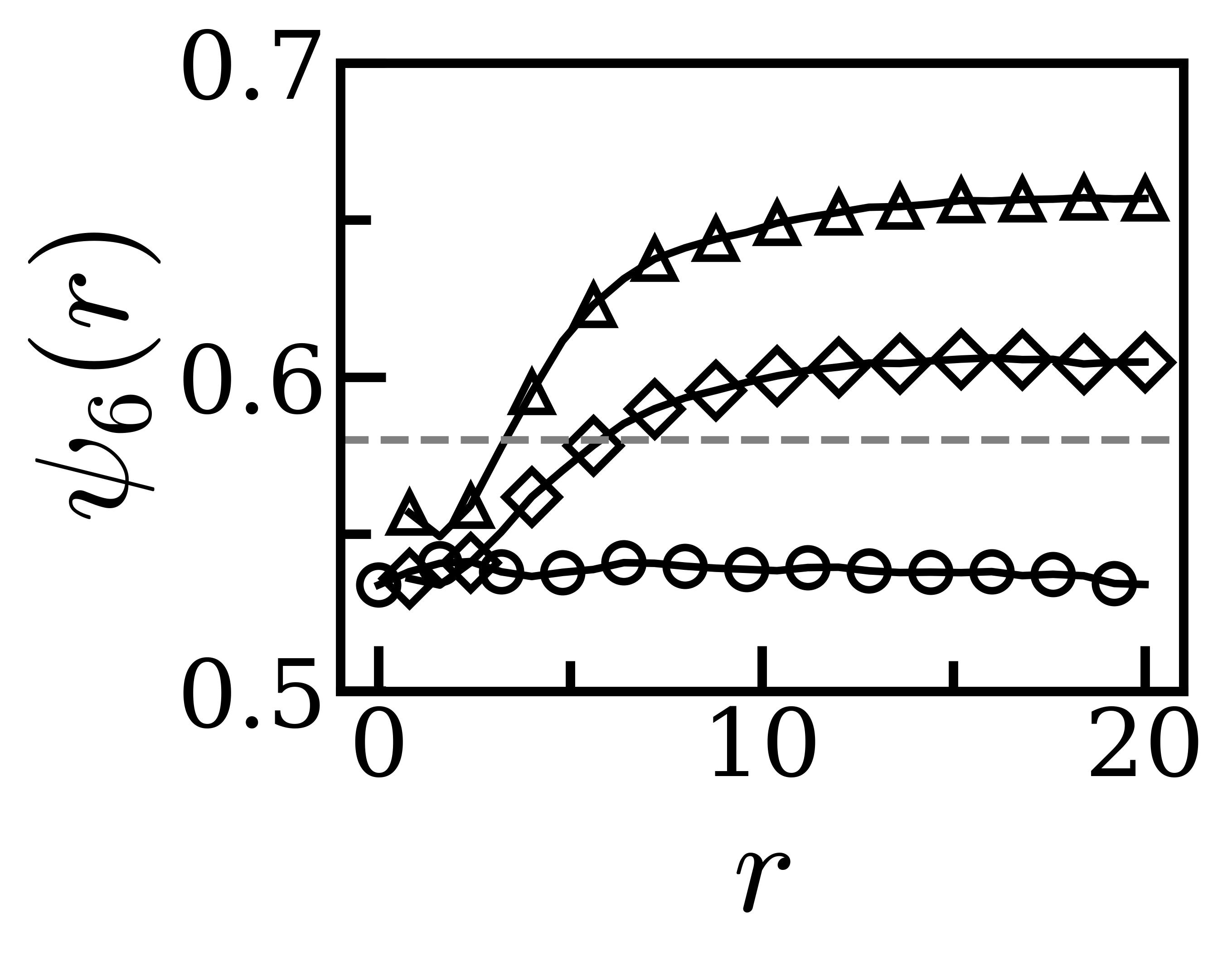}
        \put(-95,85){\textbf{(b)}}\\
    \hspace{0 cm}\includegraphics[width=0.3\linewidth]{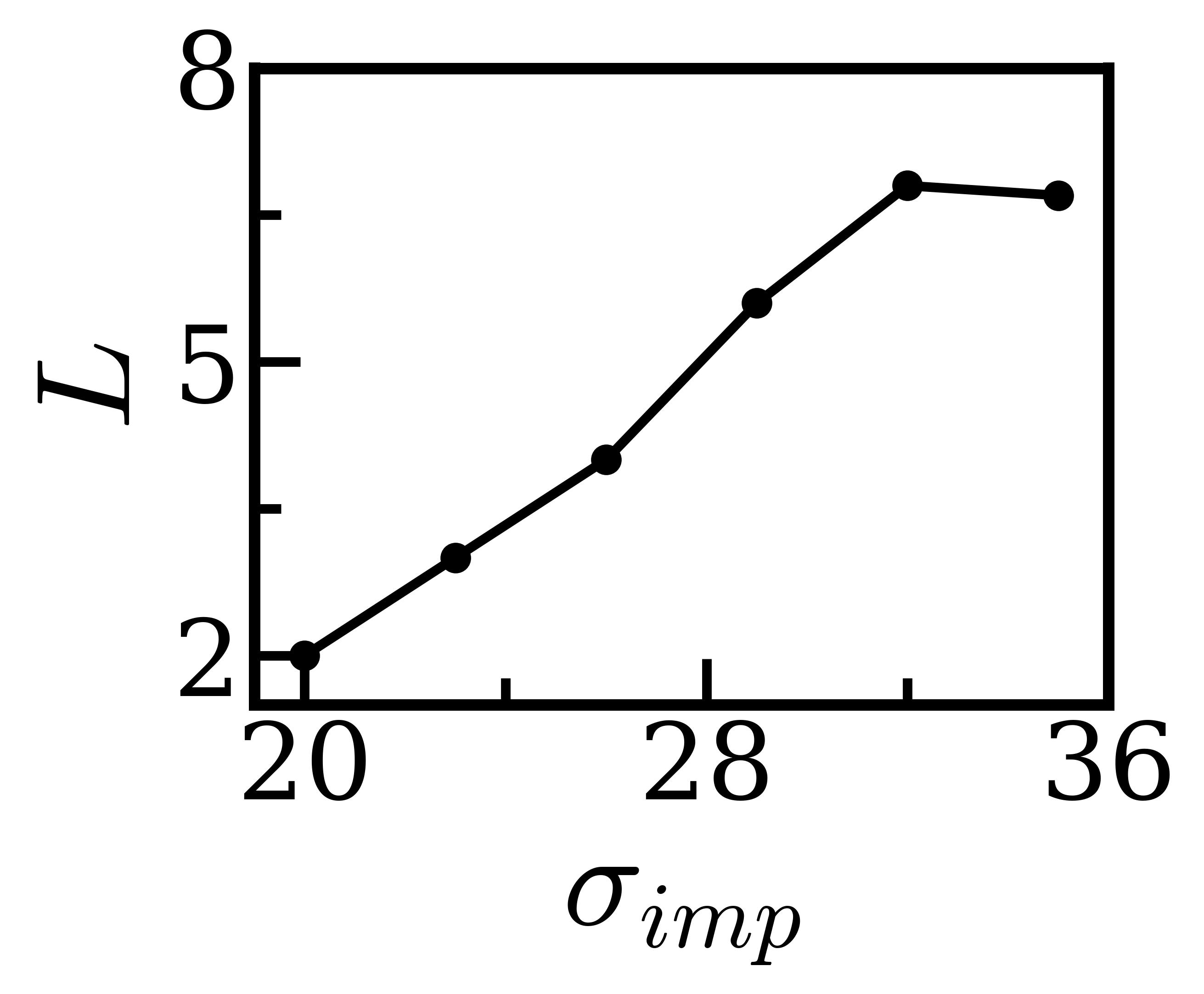}
        \put(-95,85){\textbf{(c)}}
    \includegraphics[width=0.34\linewidth]{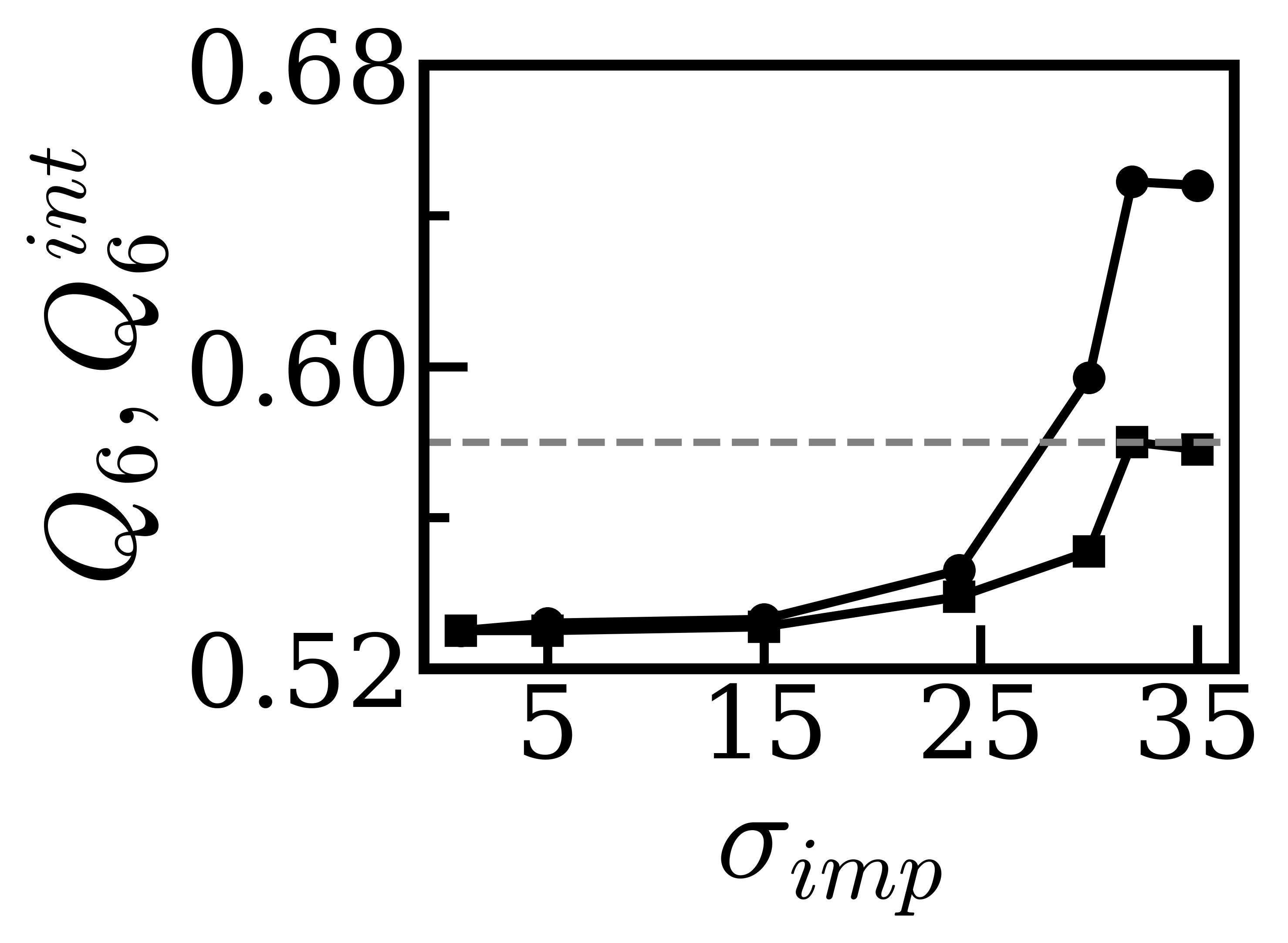}
        \put(-95,85){\textbf{(d)}}\\

    \caption{ \small{(a) Density profile around the impurity of size $\sigma_{imp}=32$. \textbf{(b)} Local bond order parameter$\psi_6(r)$ versus $r$ plots for $\sigma_{imp}$=5 (circle), $\sigma_{imp}$=30(diamond) and $\sigma_{imp}$=32 (triangle). \textbf{(c)} Interfacial width (L) vs $\sigma_{imp}$ plot. \textbf{(d)} Total bond order parameter, $Q_6$ vs $\sigma_{imp}$. Filled circles represents $Q_6$ away from impurity and squares are represents $Q_6$ within the interfacial width.}}

\end{figure}

We examine the local bond order in the  system through the distance($r$) dependent bond order parameter $\psi_{l}(r)$  ($l=5,6,7$), averaged over all the particles within a shell between distances $r$ and $ r+dr$ from the surface of the impurity particle ($r$ =0) and the equilibrium configurations. This quantity describes the ordering of the colloidal particles at various distances from the impurity particle surface. We show $\psi_6(r)$ versus $r$ plot in Figure 2(b).  We observe that $\psi_6(r)$ is close to the  bulk fluid like value ($Q_6$=0.53)  for all $r$ for low values of $\sigma_{imp}$, implying no perturbation of the fluid structure for small impurities. For larger $\sigma_{imp}$, $\psi_6(r)\simeq Q_f$ (=0.53) near the surface and saturates to values exceeding $Q_f$ beyond an interface of width $L$. We define the saturated value of 6-fold bond order parameter far from the impurity surface as $Q_6$. $Q_6$ increases with $\sigma_{imp}$ as shown in  Figure 2(d). 

We define the interfacial region up to the distance  where change of $\psi_6(r)$ attains a value half of $Q_6$. We show in Figure 2(c) $L$ versus $\sigma_{imp}$ plot. We observe that $L$ increases linearly with $\sigma_{imp}$ and then saturates around $L\simeq$ 7 for $\sigma_{imp}\geq 32$. We define the interfacial bond order parameter $Q^{int}_{6}$ by integrating over $\psi_6(r)$ within the interfacial width $L$ for each $\sigma_{imp}(\geq 20)$. Since for lower $\sigma_{imp}$ there is no pronounced variation in $\psi_6(r)$, we take $Q^{int}_{6}=\psi_6(r=0)$.We show in Figure 2(d), $Q^{int}_{6}$ versus $\sigma_{imp}$ data as well. $Q^{int}_{6}$ increases slightly and then saturates, the values remaining as in the bulk fluid phase. Thus, the interfacial particles are orientationally disordered despite having high particle density. Thus, the impurity surface is wet by an orientationally disordered fluid phase of high density.

We show $\psi_5(r)$ and $\psi_7(r)$ also in the presence of the impurity particle in Figure 3(a) and 3(b). The number of five and seven-fold coordinated particles near the surface of impurity is as high as those in the bulk fluid phase and they do not decay much with distance from the impurity surface. This trend is not significantly dependent on $\sigma_{imp}$. Thus, defects sites are stabilized at the impurity surface even for large $\sigma_{imp}$. Such defects proliferate far away from the impurity surface.

2D systems often shows hexatic phase with long ranged bond-orientation order but no long ranged translational order due to presence of defect sites, intervening the crystalline and the fluid phase\cite{r60,r59}. The presence of defects in our system makes us to check if the ordering in 6-fold coordination is due to hexatic phase. We calculate to this end the bond orientation correlation function (Eq.5, Methods) $G_6(r)$. We show in Figure 3(c) the $G_6(r)$ vs $r$ plots in the presence of impurity.  $G_6(r)$ decays exponentially with $r$ which indicates that the system is in fluid phase with short ranged bond-orientation function for small $\sigma_{imp}$. Further, with increasing $\sigma_{imp}$ the decay changes to an algebraic dependence,$G_6(r) \propto r^{-\zeta}$, suggesting a quasi-long ranged ordered (QLRO) phase. The exponent $\zeta$ depends on $\sigma_{imp}$ as shown in  Figure 3(d), the exponent falling to very low values for large $\sigma_{imp}$. 

\begin{figure}[htbp]
\centering
    \hspace{0.2cm}\includegraphics[width=0.3\linewidth]{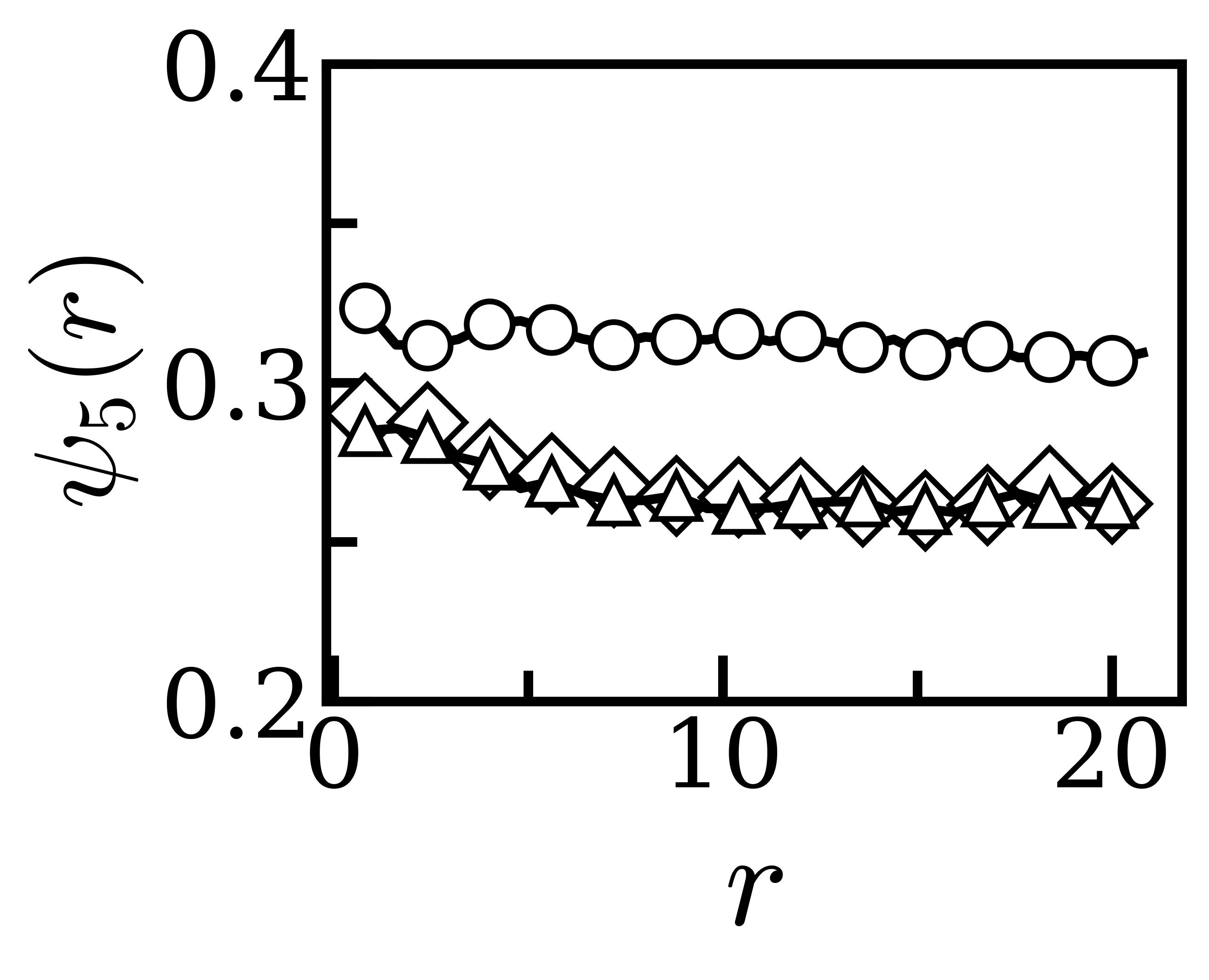}
        \put(-22,80){\textbf{(a)}}
    \hspace{0.6cm}\includegraphics[width=0.3\linewidth]{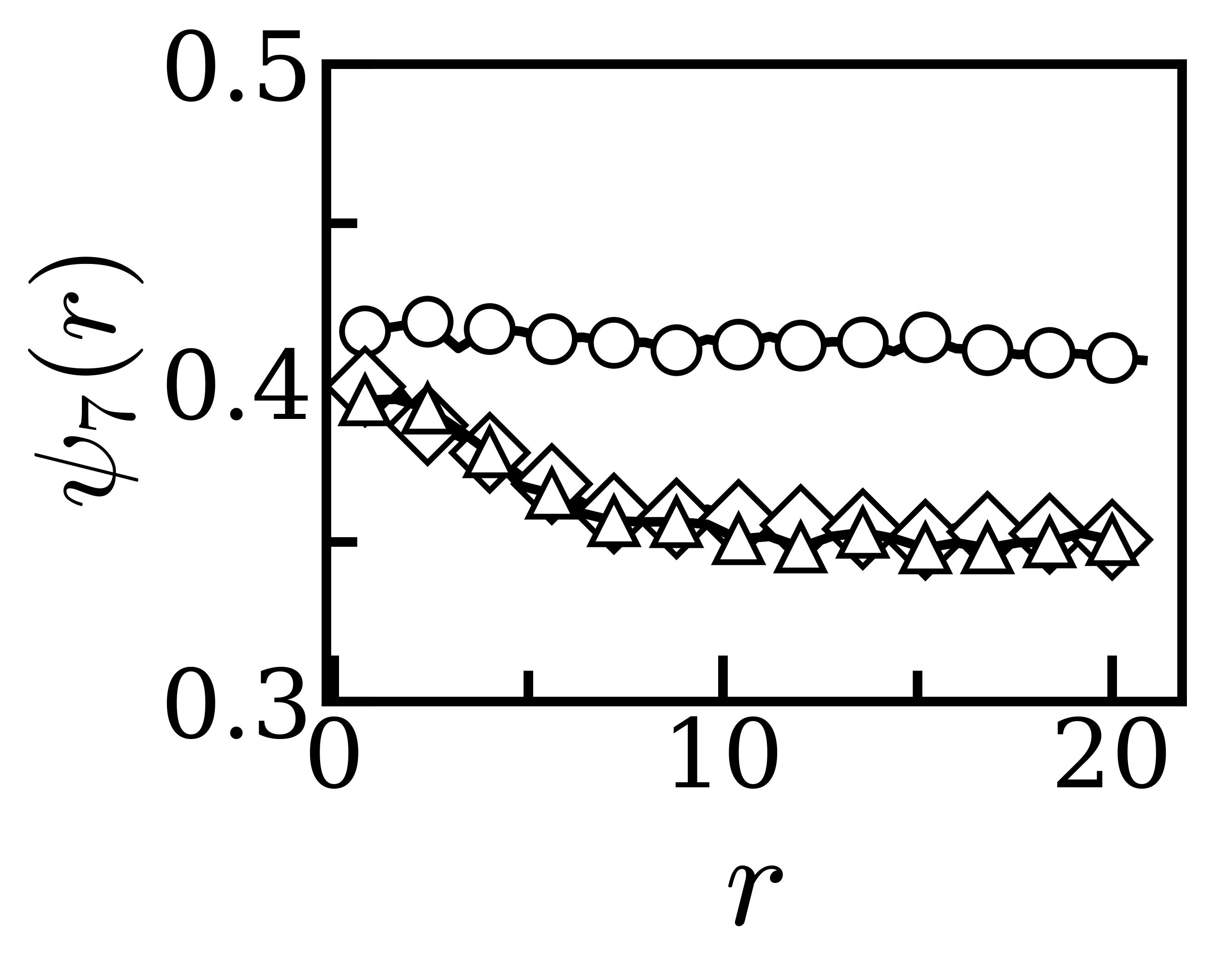}
        \put(-22,80){\textbf{(b)}}\\
    \includegraphics[width=0.34\linewidth]{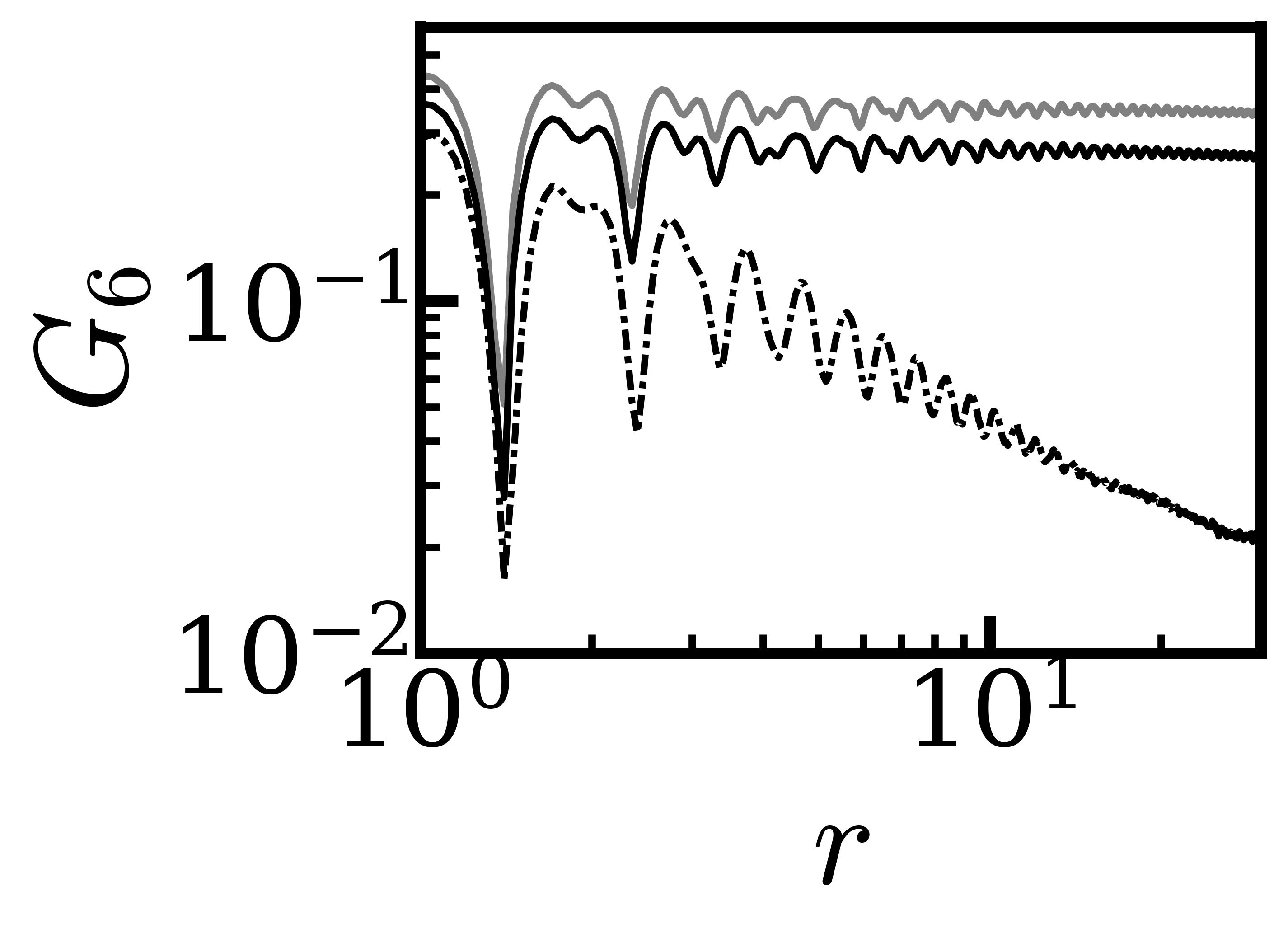}
        \put(-22,78){\textbf{(c)}} 
    \includegraphics[width=0.34\linewidth]{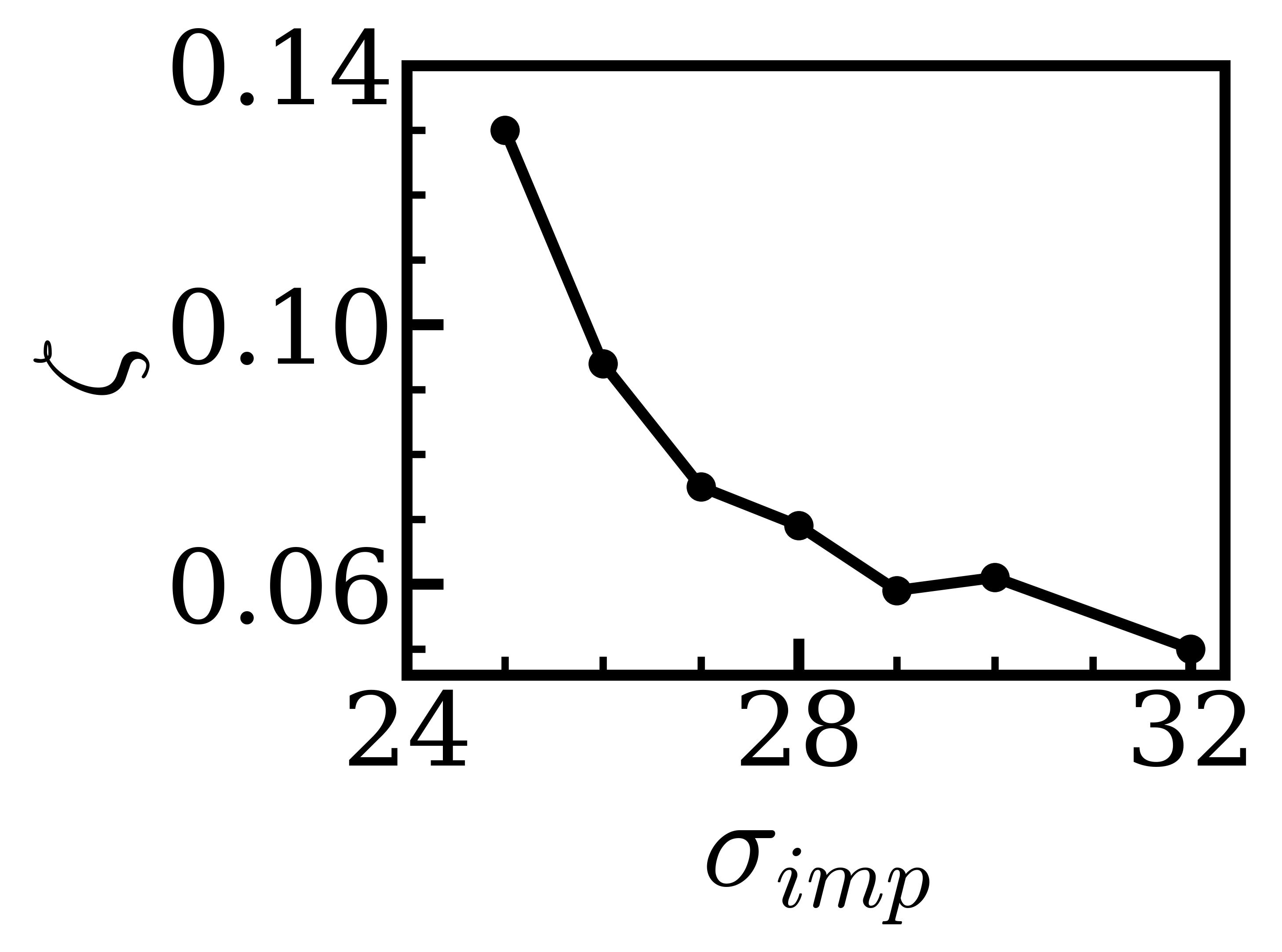}
        \put(-27,80){\textbf{(d)}}
        
    \caption{\small{\textbf{(a)} $\psi_5$ and \textbf{(b)} $\psi_7 $ vs r plot for different impurity sizes $\sigma_{imp}$=5 (circle),$\sigma_{imp}$=30 (diamond) and $\sigma_{imp}$=32 (triangle). \textbf{(c)} Spatial(r) dependence of bond orientation correlation function ($G_6$) for $\sigma_{imp}$ = 5(dashed dot),   30(solid black line) and 32(solid gray line) in log-log scale. \textbf{(d)} $\zeta$ vs $\sigma_{imp}$ plot.}} 

\end{figure}

The power law decay implies that the system can be either in a hexatic phase or there is  phase co-existence between fluid and hexagonal crystalline phases\cite{r35}. We compute the bond orientation susceptibility to check this. We divide the system into smaller boxes of different sizes $L_{b}$. We compute $\chi_6$ [Eq. 6, Methods] for each subsystem at every configuration, excluding the region occupied by the impurity particle in various subsystems. Then we construct the distribution of $\chi_6$,$P(\chi_6)$ considering each of the blocks over all the independent trajectories. In the pure solid, liquid and hexatic phase,  $P(\chi_6)$ is single peaked\cite{r36}.

Figure 4(a) shows $P(\chi_6)$ for different $L_b $ for  $\sigma_{imp}=$ 32 where orientation correlation function decays algebraically very slowly ($\zeta=0.05$). Here we also observe that there are two peaks for all box sizes, suggesting that the QLRO phase is a co-existence between a fluid and crystalline solid phase in the system\cite{r35} rather than a hexatic phase. 
In order to highlight the phase coexistence in QLRO, we identify solid and liquid like particles in the system. A particle $i$ is called solid-like if it satisfies  $\psi_6(i)\psi_6(j)$>0.36 with its neighbor particle j and connected with more than 5 particle, else it is liquid-like. A map over liquid and solid like particles in an equilibrated configuration for $\sigma_{imp}$ has been shown in Figure 4(b). The particles surrounding the impurity are primarily liquid, while there are pockets of liquid patches distributed even in the background of the solid particles. 

\begin{figure}[H]
\centering
    \includegraphics[width=0.33\linewidth]{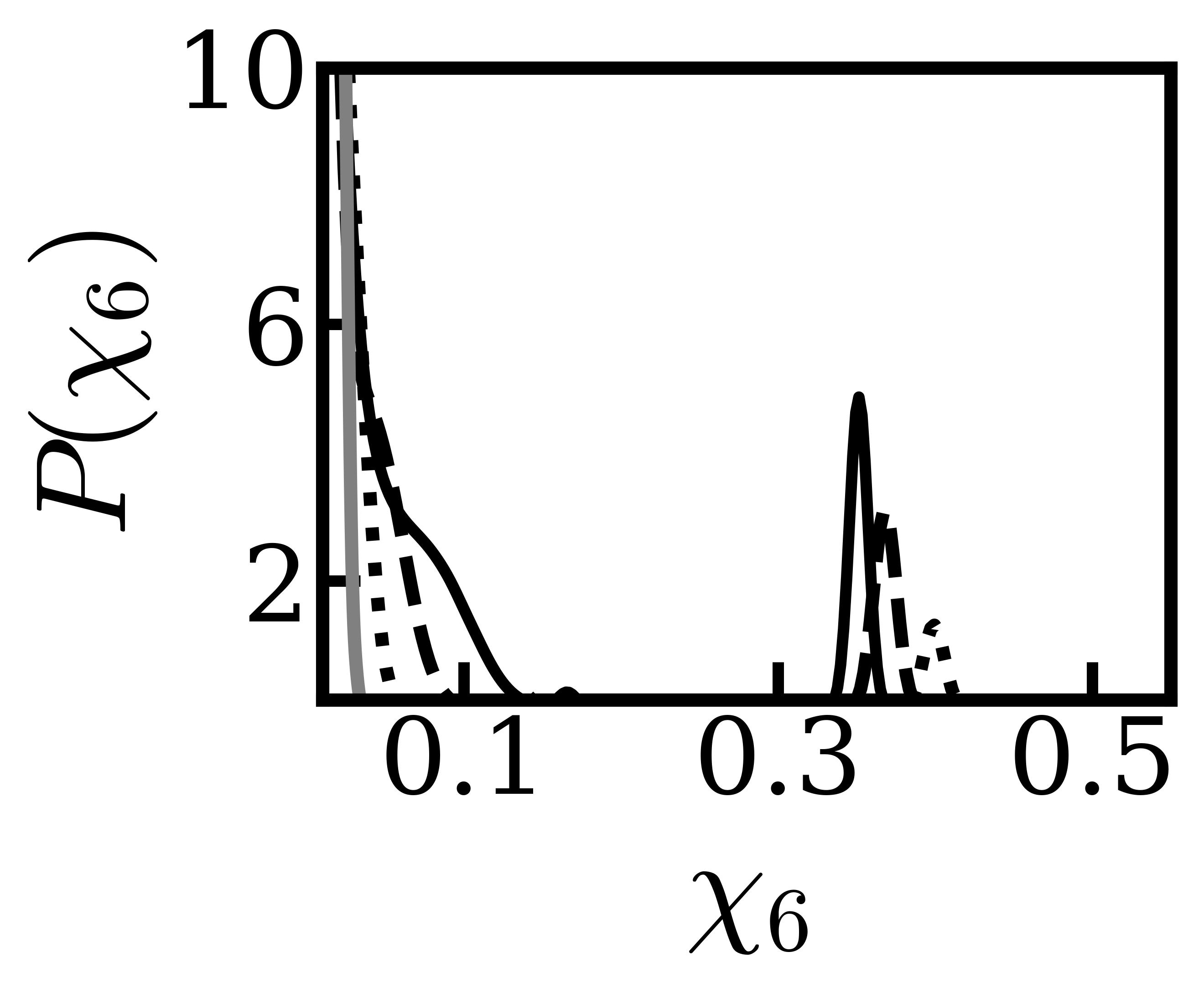}
        \put(-100,95){\textbf{(a)}} \\
    \hspace{1.5 cm}\includegraphics[width=0.38\linewidth]{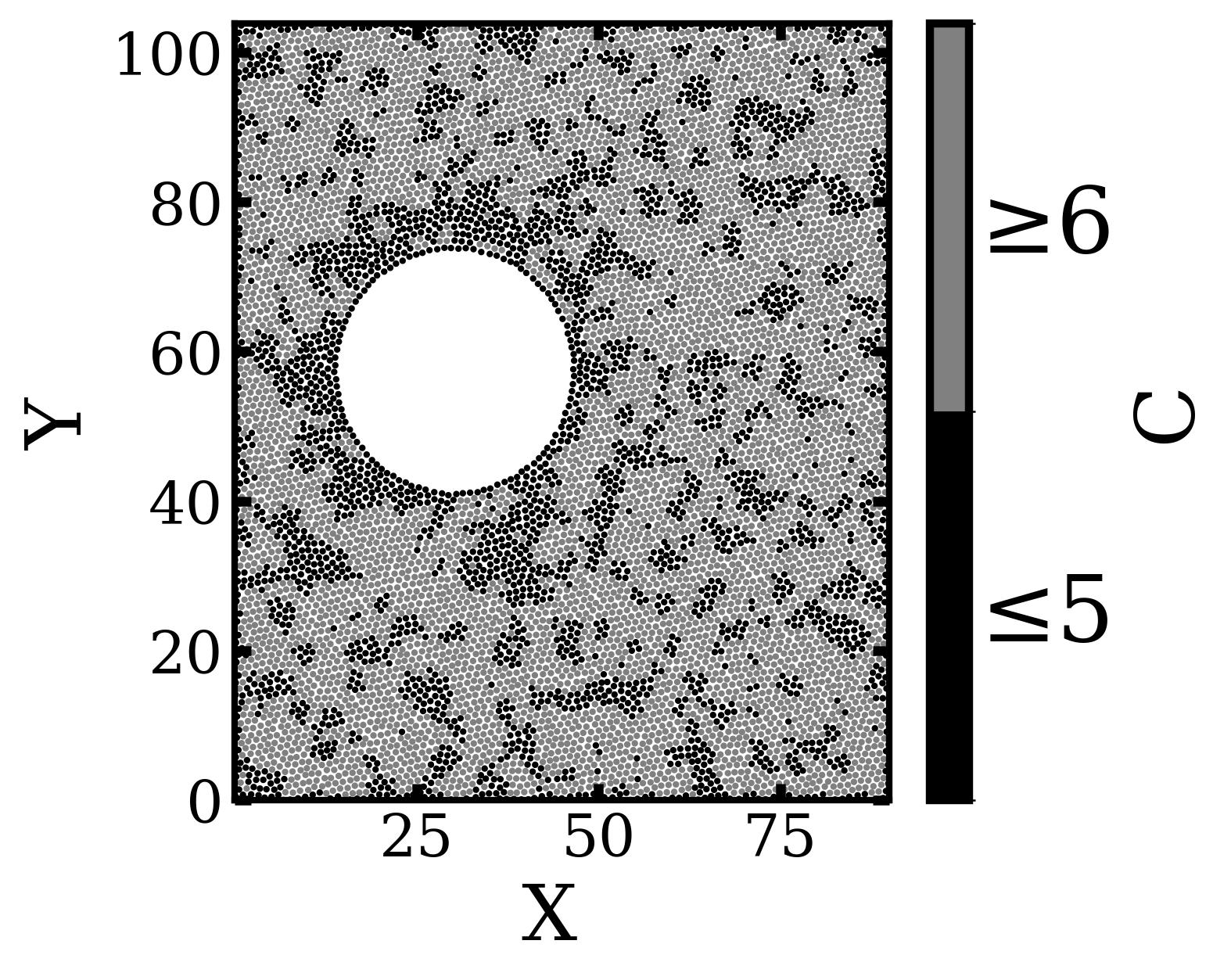}
        \put(-132,115){\textbf{(b)}}

    \caption{\small{\textbf{(a)} Probability distribution of susceptibility, $P(\chi_6)$ versus $\chi_{6}$ plots for $L_b=$ L/4 (gray solid line), L/8(dashed dot), L/16(dashed) and L/32(solid line) subdivisions of the box. \textbf{(b)} Map of solid-like and liquid- like particles shown over an equilibrium configuration. The black ones  are liquid-like and gray ones ones have solid-like connections(C). }} 

\end{figure}
Further we calculate solid fraction $\Theta_c$, defined as the ratio fraction of solid like particles.  Figure 5(a) shows that $\Theta_c$ increases continuously with $\sigma_{imp}$ and nearly 60$\%$ particles are solid-like for $\sigma_{imp}\geq 32$. It may be noted that nearly 50$\%$ growth in $\Theta_c$ takes place around $\sigma_i$=24. The $Q_6$ exceeds $Q_f$ around similar value of $\sigma_{imp}$ as can be seen in Figure 2(d) by the dashed line. The $C_V$ data in presence of impurity in Figure 5(b) shows peak at $\sigma_i$=24, suggestive of transition from fluid to QLRO. The $Q_6^{int}$ data show that the interfacial particles remain liquid like despite high local particle density $\rho(r)$ (Figure2(a)) across the transition, although $Q_6^{int}$ increases with $\sigma_{imp}$. This is as well observed in the particle map in Figure2(b).

In the presence of impurity the effective area available for the colloidal particles decreases. We check if this helps in the freezing of the system in presence of impurity. We run simulations with bulk colloidal particles without any impurity, but keeping the area  of the bulk system the same that in the case of presence of a single quenched impurity. We consider the case with  $\sigma_{imp} =32$.  We observe  $Q_6\simeq$ 0.53. Thus, the impurity-host interaction plays a crucial role to induce orientational order in the system. In order to make the role of the impurity-host interaction more apparent, we compute the pressure $P$ of the entire system using the formula used in Ref.\cite{thompson2009}. $P$, shown in Figure 5(c), increases with $\sigma_{imp}$. The enhanced pressure induce the QLRO.

\begin{figure}[htbp]
\centering
    \hspace{0.2cm}\includegraphics[height=4.1cm]{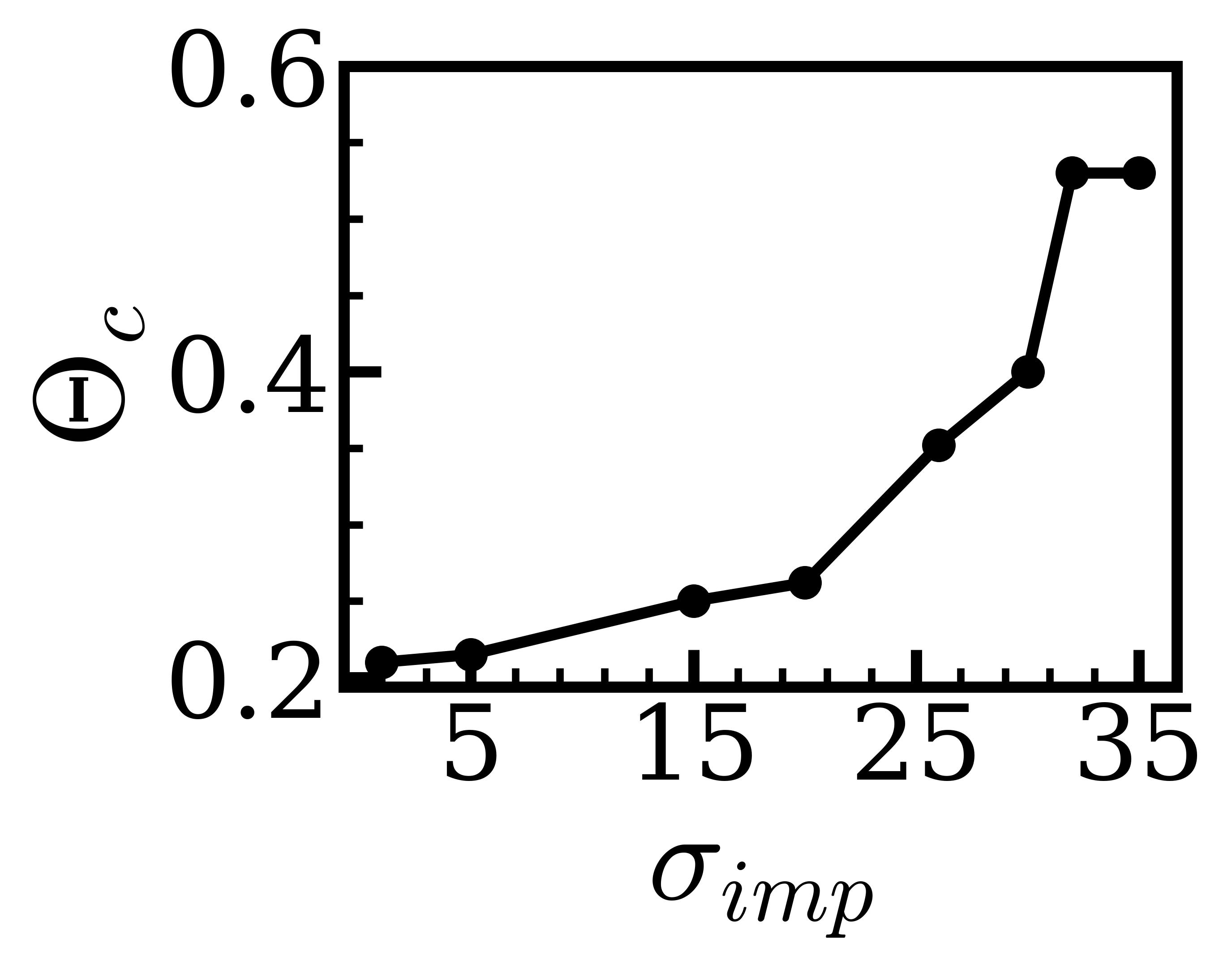}
        \put(-100,95){\textbf{(a)}} \\
    \includegraphics[height=4.0cm]{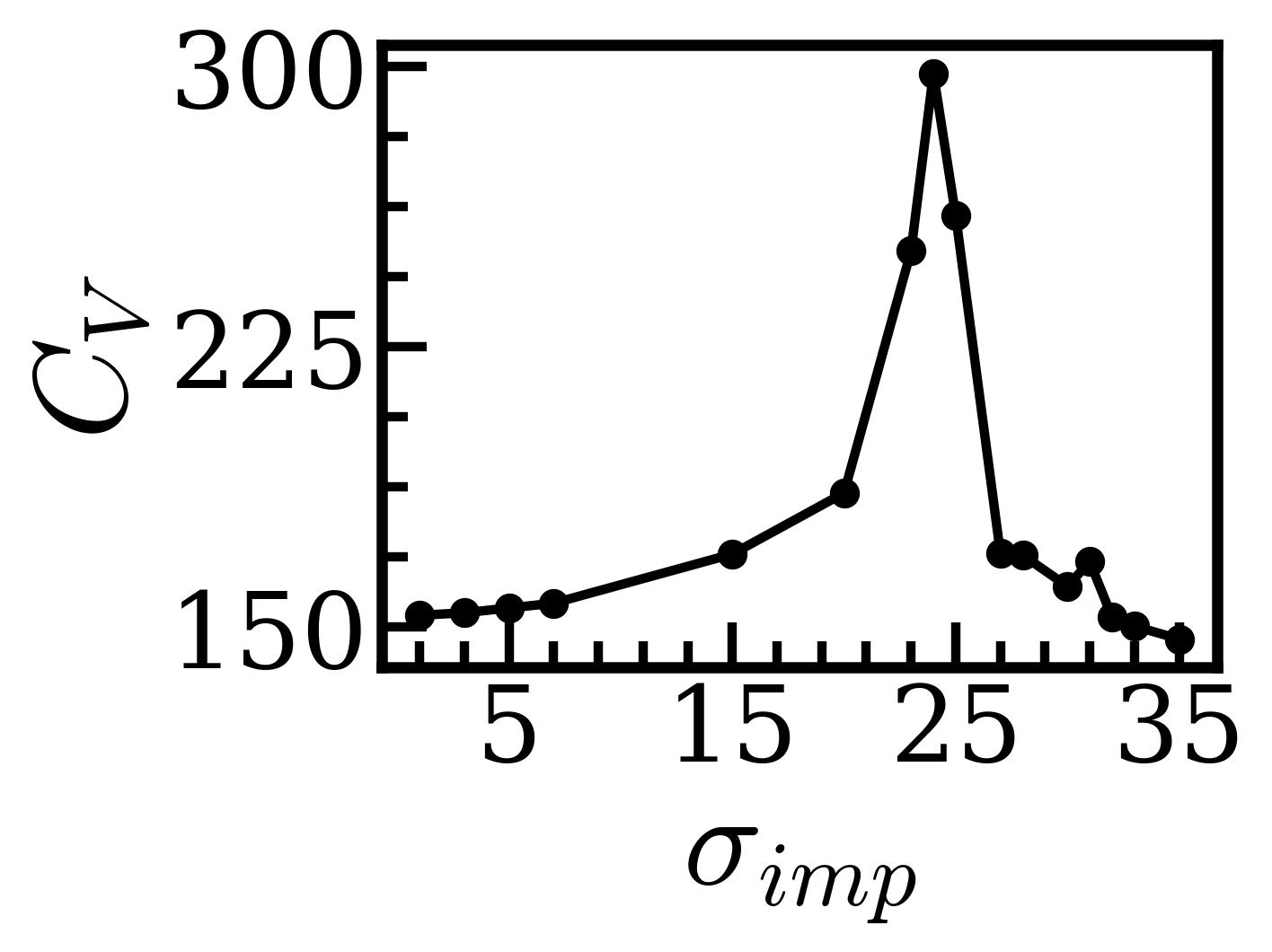}
        \put(-100,95){\textbf{(b)}}\\
    \hspace{0.4 cm}\includegraphics[height=4.1cm]{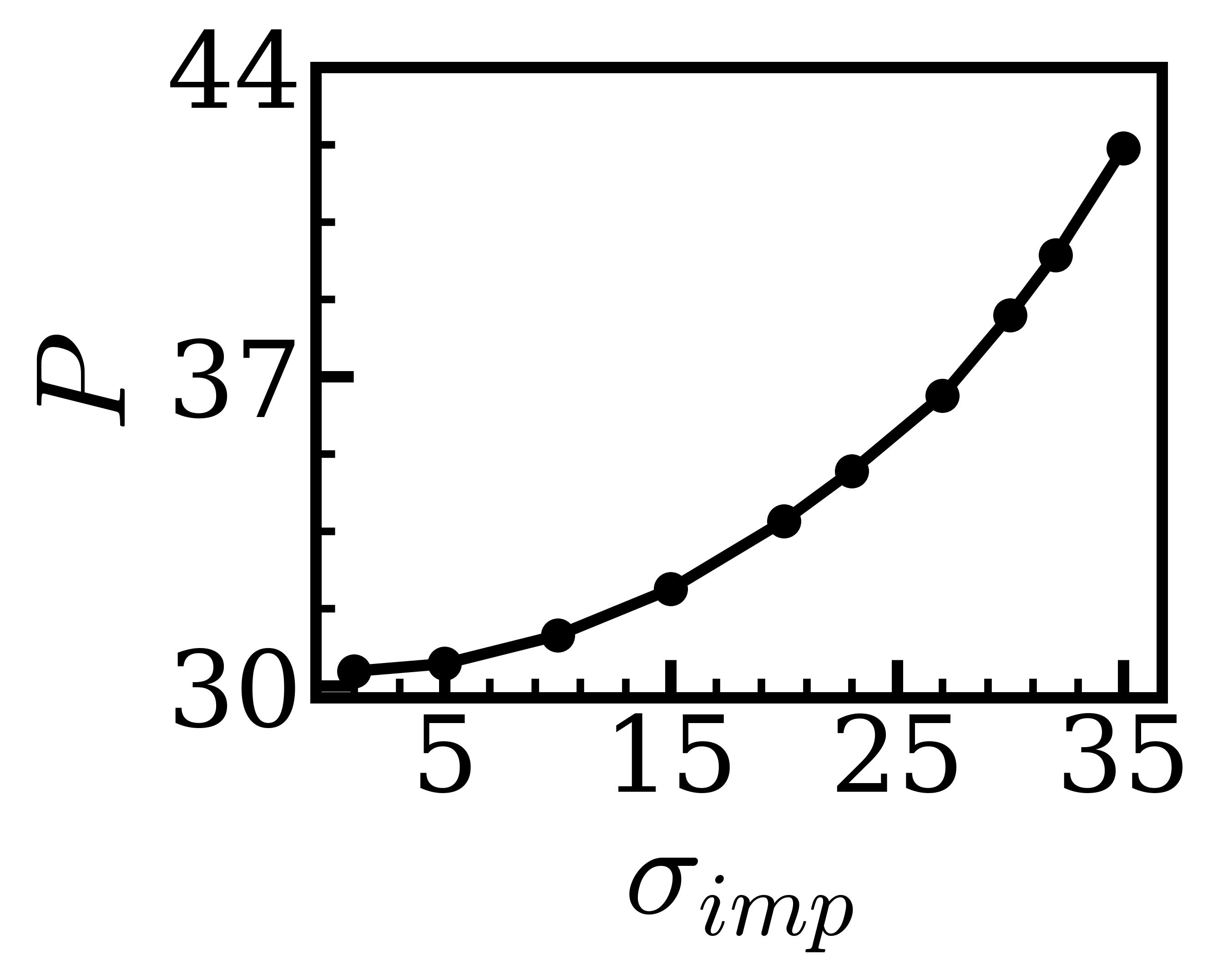}
        \put(-100,95){\textbf{(c)}}

    \caption{\small{ \textbf{(a)} Solid fraction ($\Theta_c$) vs $\sigma_{imp}$ away from the interfacial region, where $\psi_6$ saturates. \textbf{(b)} Specific heat $(C_V)$ vs $\sigma_{imp}$.  \textbf{(c)} Pressure(P) vs $\sigma_{imp}$. }}

\end{figure}
\subsection{Particle dynamics}

Next we characterize the dynamics of the system in the presence of impurity. First, we calculate mean squared displacement (MSD) of the colloidal particles, defined as $\langle(\Delta r)^2\rangle=\langle\vec r(t)- \vec r(t_0))^2\rangle$. Here the average is taken over different time origins $t_0$ over  different trajectories. 
Figure 6(a) shows that the  MSD for $\sigma_{imp}=5.0$ is linear $\langle(\Delta r)^2\rangle \sim t$ in time after a ballistic region $\langle(\Delta r)^2\rangle \sim t^2$ for very short times, suggesting a long-time diffusive motion. In the phase-coexistence region with QLRO ($\sigma_{imp}=$ 32), the MSD values are much lower than that in the fluid. The time dependence changes to linear at large time from ballistic behavior at short times via sublinear dependence, $\langle(\Delta r)^2 \rangle \sim\sqrt{t}$ at intermediate times. Thus, the particles motion is diffusive in the long time limit. The slope of the linear dependence defines the diffusion coefficient $D$ of the fluid.  Figure 6(b) shows that $D$ decreases linearly with  $\sigma_{imp}$ as the system approaches the QLRO  state till $\sigma_{imp} \sim$ 24, exactly where the specific heat peak appears after which $D$ is vanishingly small in the QLRO. 

\begin{figure}[htbp]

\centering
    \includegraphics[width=0.35\linewidth]{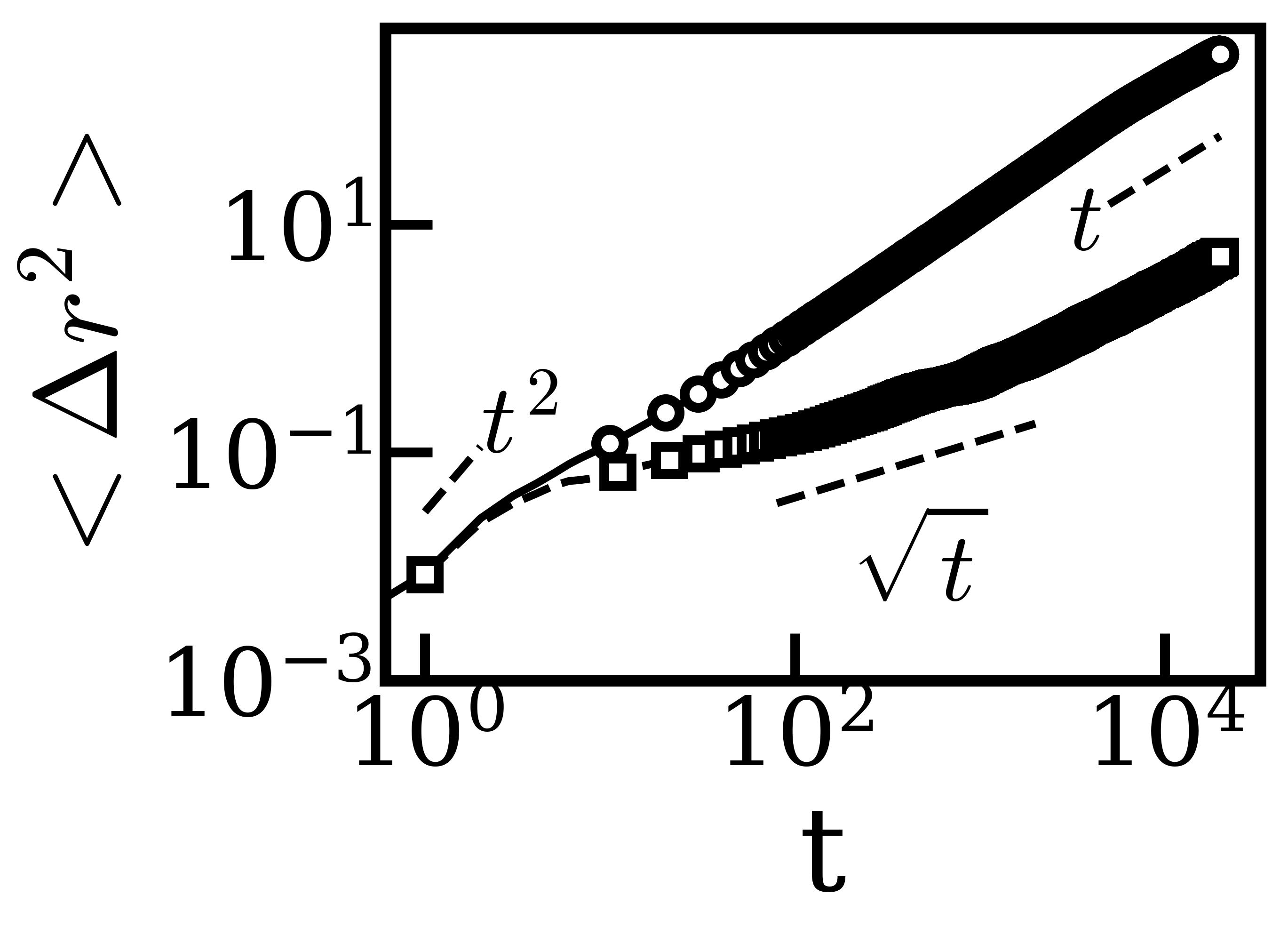}
        \put(-22,46){\textbf{(a)}} 
    \includegraphics[width=0.33\linewidth]{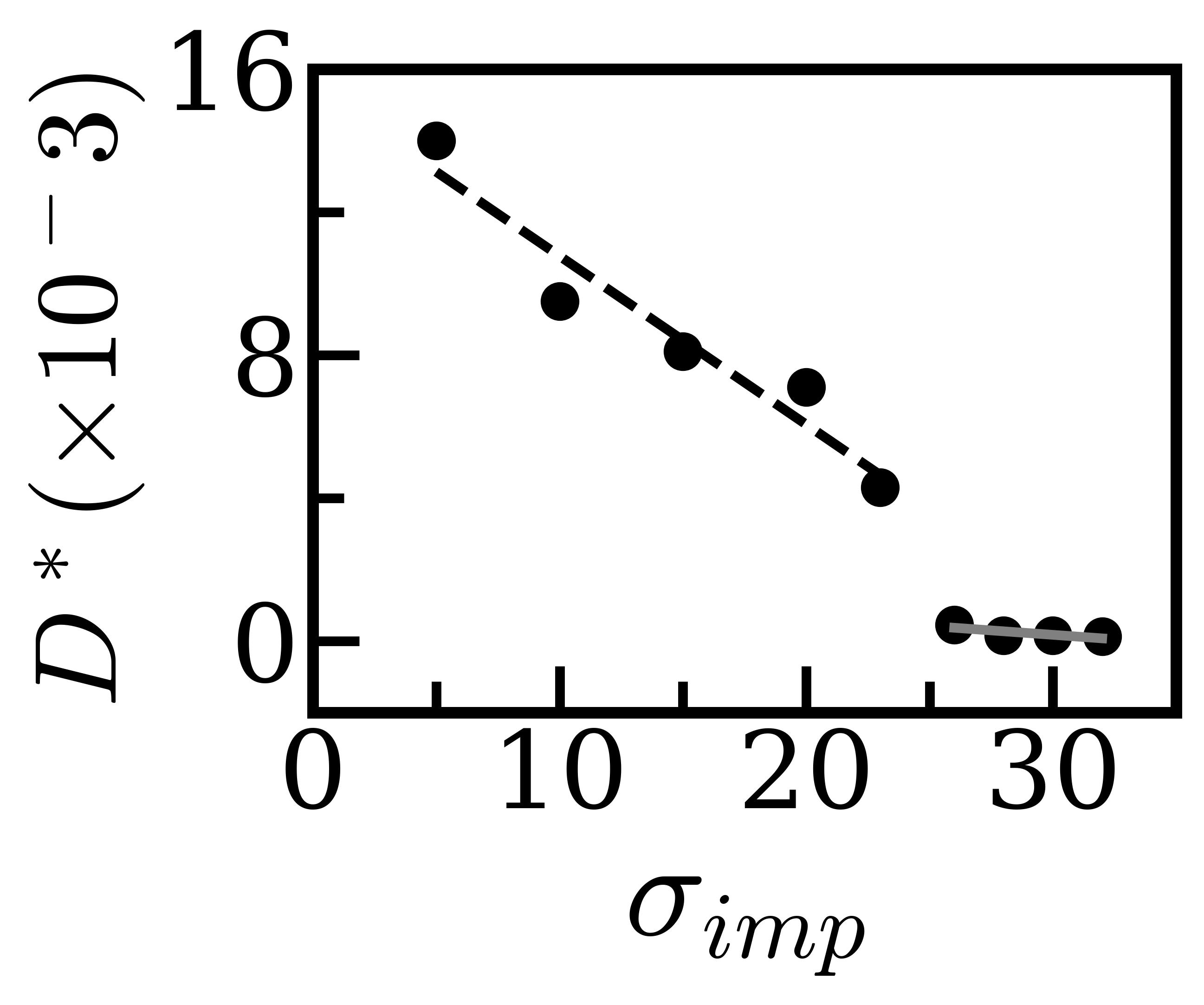}
        \put(-28,52){\textbf{(b)}}\\
    \hspace{0.5 cm}\includegraphics[width=0.35\linewidth]{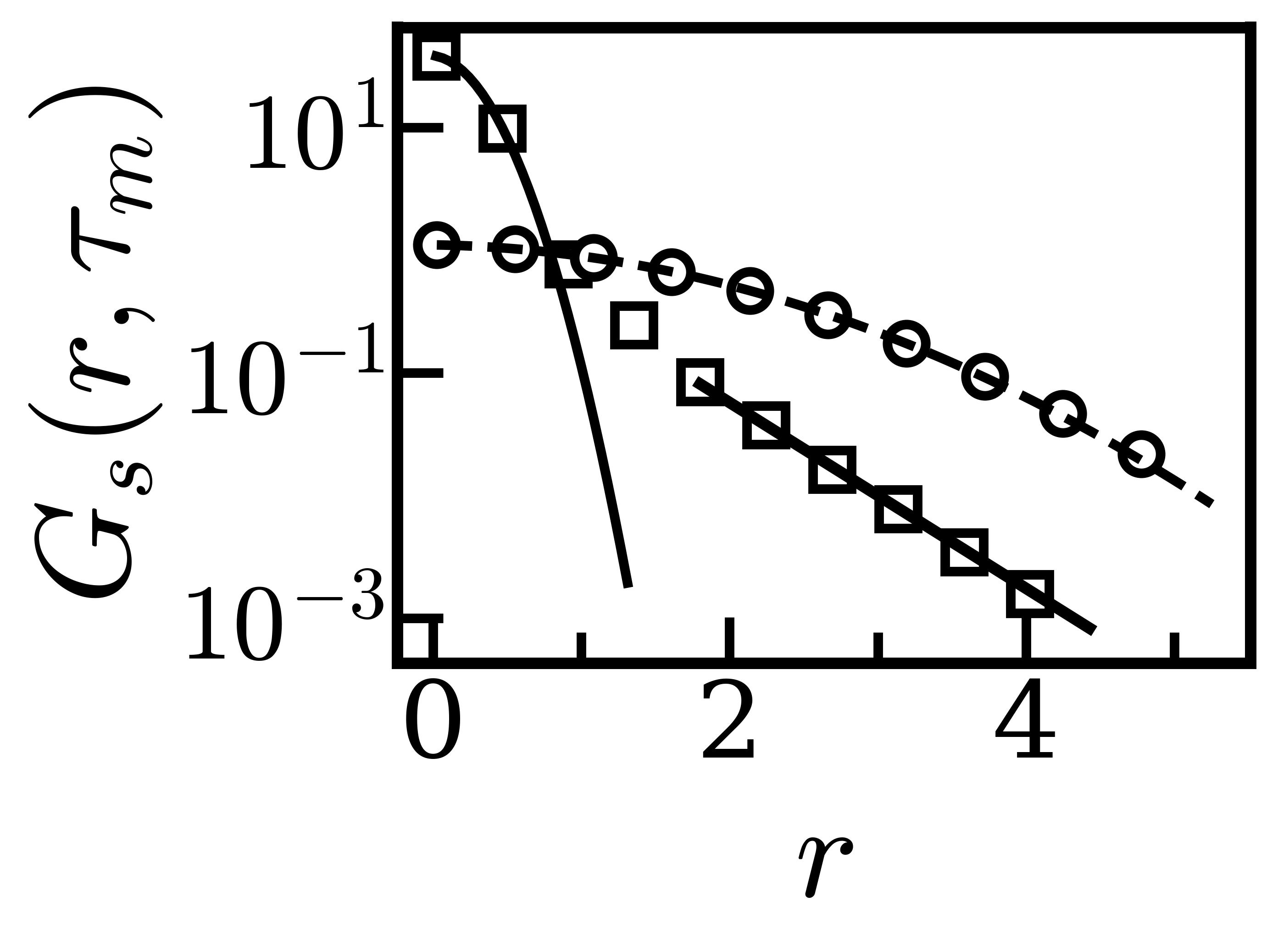}
        \put(-22,46){\textbf{(c)}}
    \includegraphics[width=0.38\linewidth]{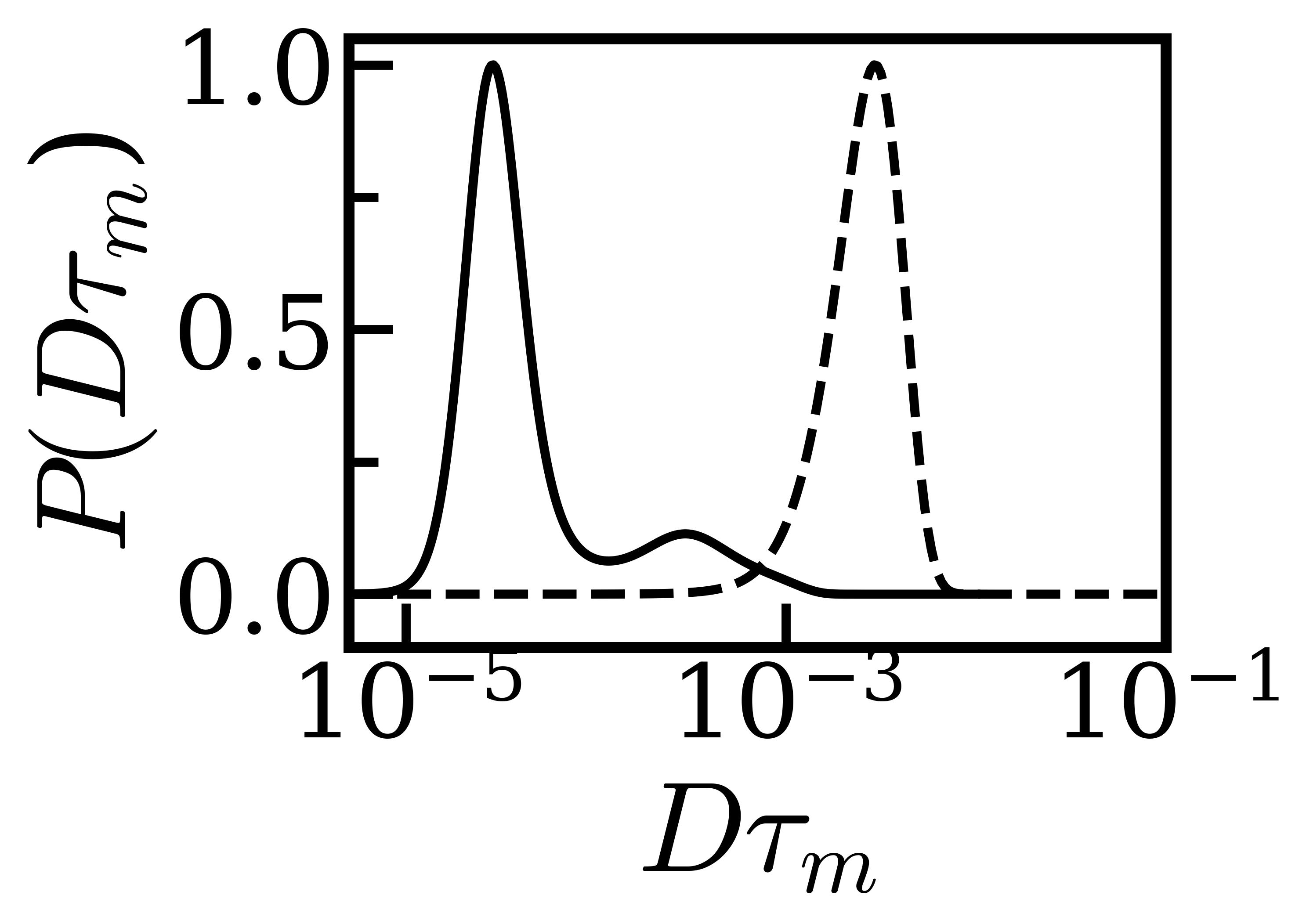}
        \put(-38,46){\textbf{(d)}}
        \put(-89,40){\vector(0,1){20}}\\
    
    \caption{\small{\textbf{(a)} MSD for different impurity sizes like $\sigma_{imp}$=5 (circle) and $\sigma_{imp}=32$(square). \textbf{(b)} Diffusion constant ($D_{t}$) vs  $\sigma_{imp}$ plot. The dashed line is linear fit of the data.  \textbf{(c)} Self part of the van Hove function for $\Delta t^* =1413\tau$ for $\sigma_i=5$(circle) and $\sigma_{imp}=32$(square). lines are used to show fitted data. \textbf{(d)} Distribution of diffusivity for $\sigma_{imp}=32$( solid line) and $\sigma_{imp}=5$ (dashed line) at $\Delta t^* =\tau_m$. The arrow indicates $D_{min} \tau_m$, the minimum after the first maximum.  }}

\end{figure}

We also explore if the heterogeneous structure in QLRO phase leads to heterogeneity in dynamics, namely, coexistence between fast and relatively slow moving particles. The particle dynamics is given by the self part of the van Hove function, $G_s(r, \Delta t)$ (Eq. 7, Methods).  $G_s(r, \Delta t)$  gives the probability distribution of magnitude of particle displacements in time interval $\Delta t$ and measurable by the neutron scattering\cite{hansen2013theory}. This probability is Gaussian width linearly proportional to $\Delta t$, the proportionality constant being the diffusion coefficient for a normal fluid, while any deviation from Gaussianity indicate deviation from normal liquid motion, observed in many complex systems\cite{r72,r73,r74,r70}.
Figure 6(c) shows $G_s(r,\tau_m)$ where $\tau_m$ is the time at which MSD becomes 0.5. For $\sigma_i$=5, $G_s(r,\tau_m)$ is Gaussian, suggesting the purely diffusive behavior of the colloidal particles. However, $G_s(r,\tau_m)$ for $\sigma_i$=32 in the QLRO phase deviates from Gaussianity at larger $r$ where an exponential tail appears, which we confirm by fitting with the following procedure explained in Ref.\cite{r71}.

The exponential tail of the self-vH suggests multiple local diffusivity of the colloidal particles\cite{wang2012brownian}. Using Lucy's iterative method \cite{r68} we construct the distribution of local diffusivity $P(D,\tau_m)$ from $G_s(r,\tau_m)$, shown in Figure 6(d). We observe a double peaked distribution for $\sigma_{imp}=32$, suggesting dynamically heterogeneous situation, some are fast diffusing and some relatively slow diffusing particles. Where for lower impurity size ($\sigma_{imp}=5$) $P(D \tau_m)$ has one peak corresponding a single diffusion constant of particles within the system.

Further we identify the fast and slow particles in the  QLRO phase following protocol in Ref. \cite{r68} We define the fast diffusing particles as those whose squared displacement is greater than or equal to $\langle \Delta r^2 \rangle=4D_{min}\tau_m$. Where $D_{min}\tau_m$ is the minimum between first two maxima in the distribution of diffusivity curve. We show the slow and fast particles in Figure 7(a). Fast particles predominantly located in the interfacial region which is in liquid phase as shown in Figure 4(b). Figure 7(b)shows the trajectory of a particle located near the impurity surface. The particle undergoes displacements along the impurity perimeter which is facilitated due to the symmetry of the impurity-host interaction. This motion disrupts the 6-fold bond order. The particles in the region away from the interface are typically slow which is consistent with higher $Q_6^{sat}$ values. On the other hand, disordered patches in this region are caged by the slow solid like particles. A particle situated far from the impurity exhibits localized motion, shown in Figure 7(c). This reflects different dynamical modes coexisting in the QLRO phase, the interfacial dynamics being qualitatively different, compared to that in the bulk.

\begin{figure}[htbp]
\centering
    \includegraphics[height=4.6cm]{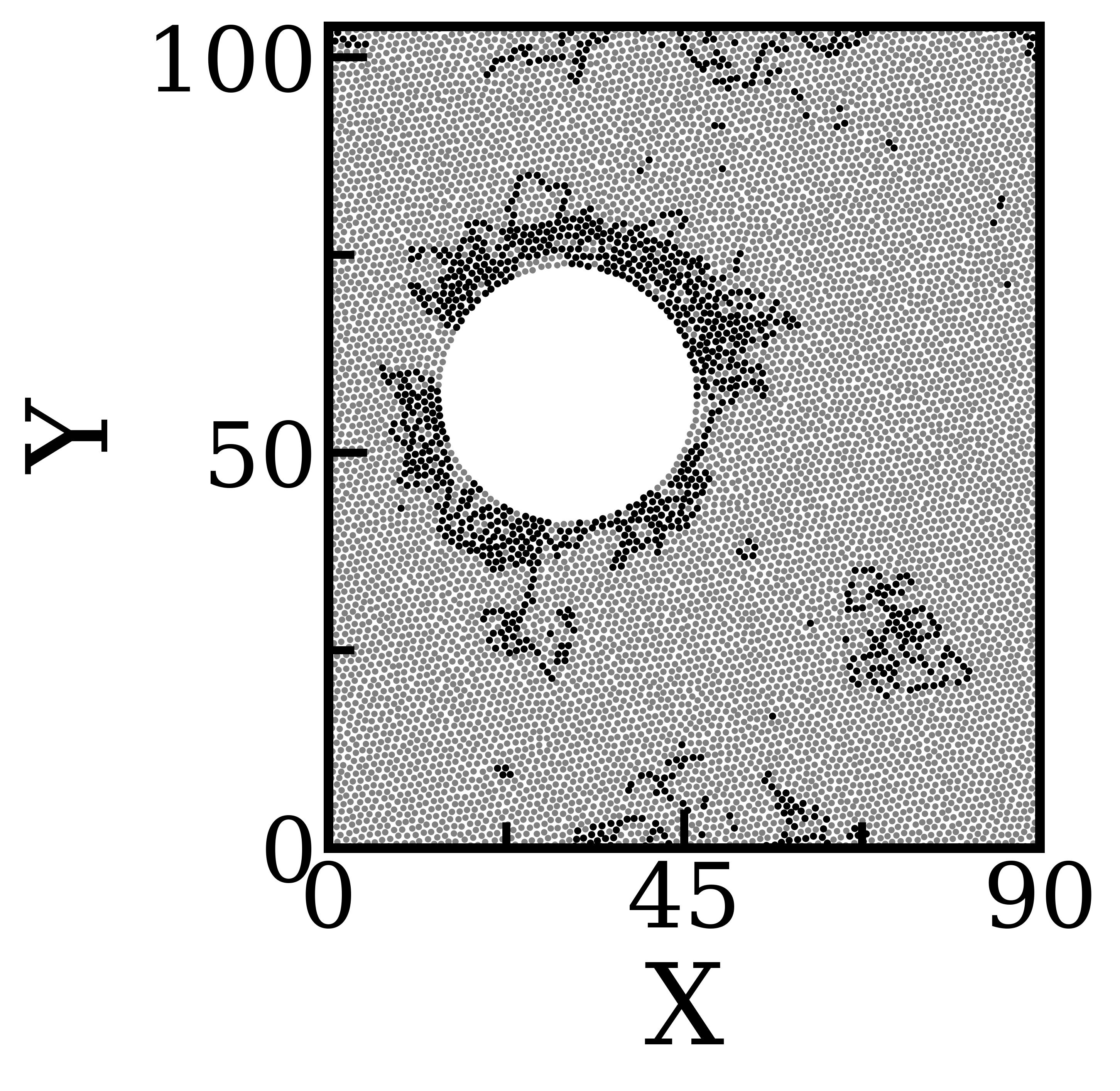}
        \put(-90,115){\textbf{(a)}}\\

    \includegraphics[height=4.8cm]{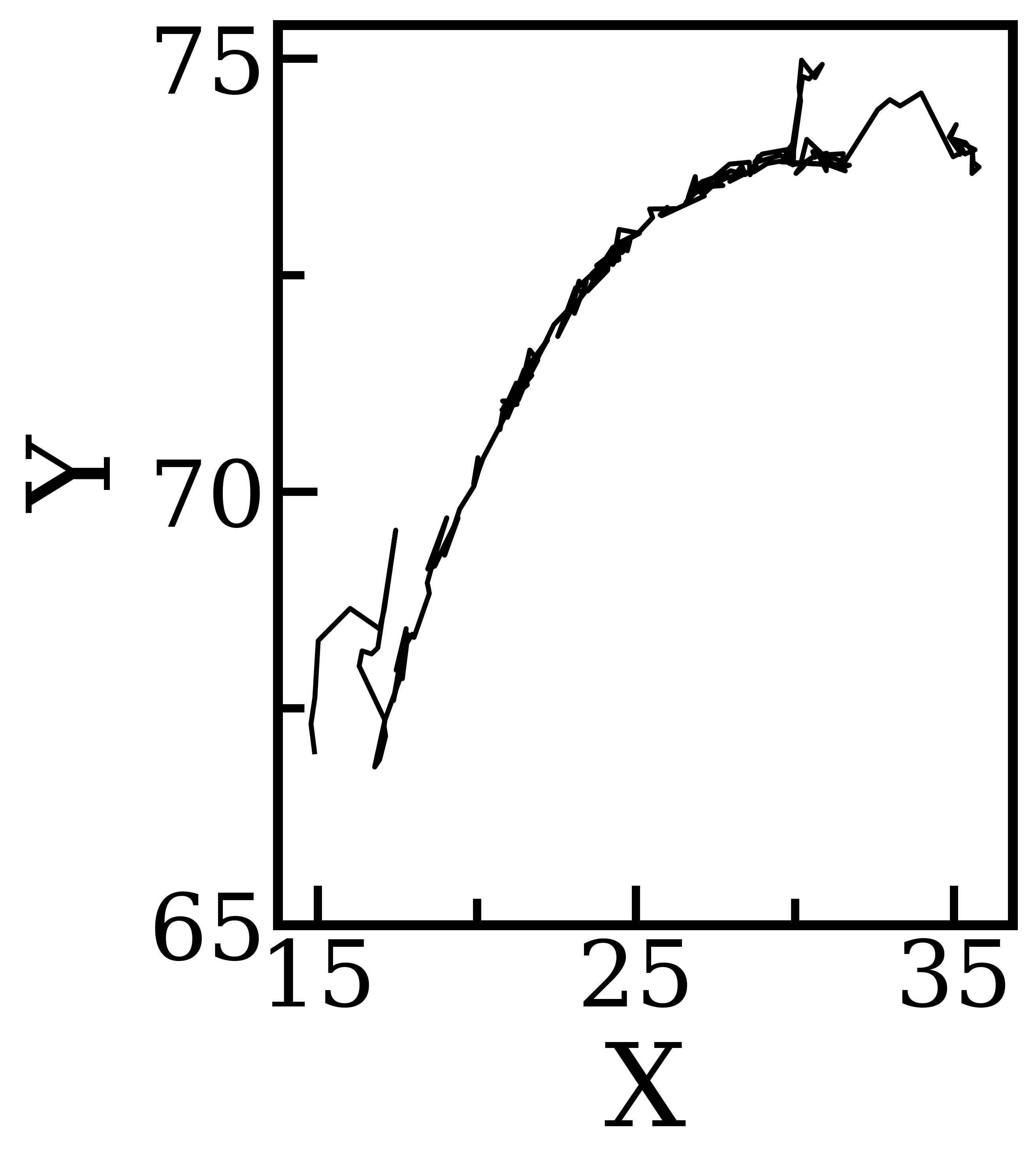}
        \put(-85,115){\textbf{(b)}}\\

    \includegraphics[height=4.8cm]{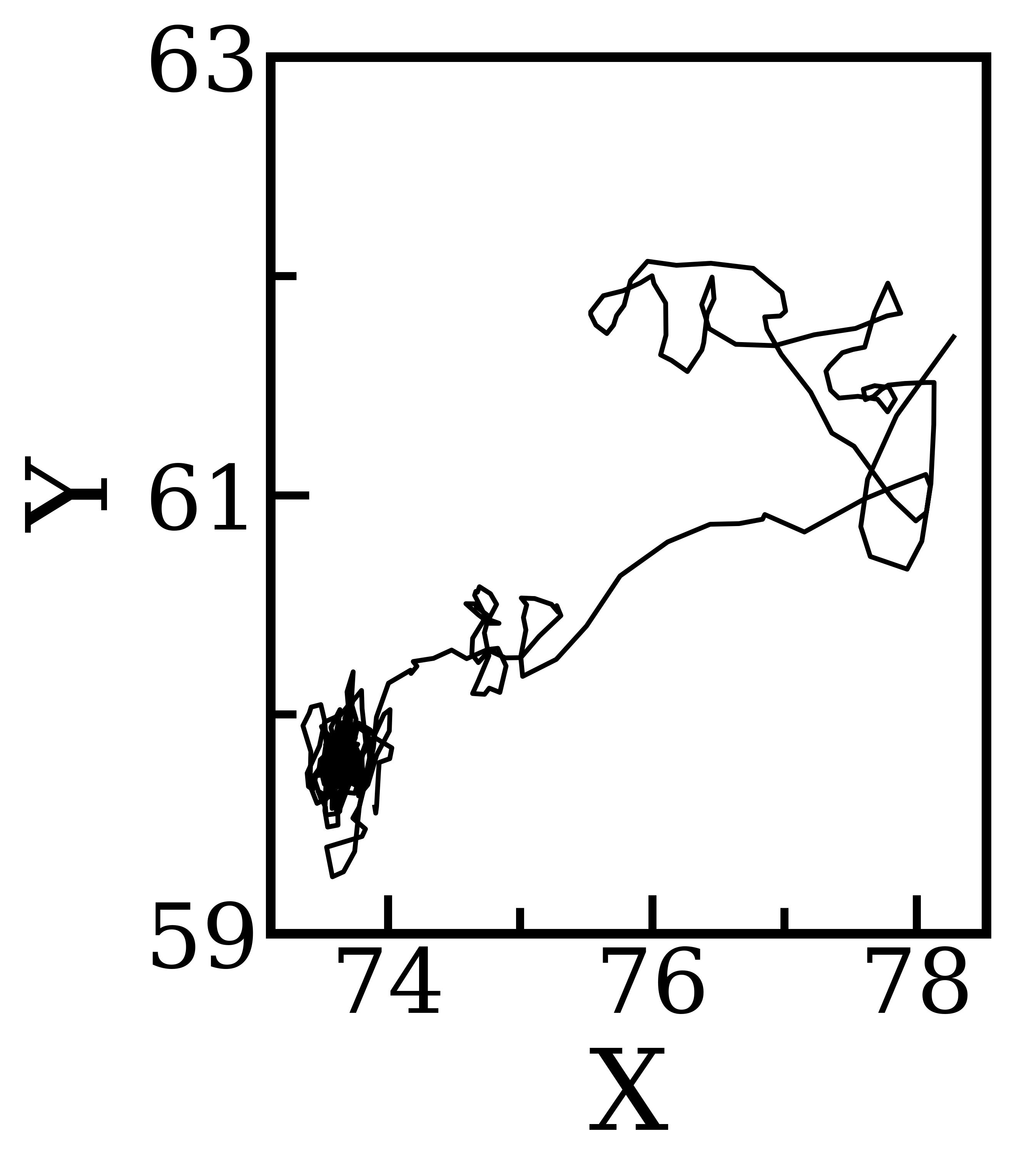}
        \put(-85,115){\textbf{(c)}}

    \caption{\small{\textbf{(a)} Slow and fast particle shown over the same equilibrium configuration in Figure 4(b). The black bullets are fast particles and gray ones are slow particle \textbf{(b)} A particle trajectory( data collected every 100 frames after equilibration) near the surface of a larger impurity ($\sigma_{imp}=32$) and \textbf{(c)} far from the impurity} }

\end{figure}


\subsection{Double impurity}

We further examine how the presence of another impurity in the vicinity of a large impurity affects the QLRO. We consider $\sigma_{imp}=32$ where an impurity particle is placed at the center of the box and a second particle of the same size pinned at various surface to surface distances ($R$). The total number of particles in the system is kept unchanged after insertion of the impurity particles. We choose the second site at six different angles (60$^0$ and multiples) about the x-axis. All the data are averaged over these six cases for a given $R$. We calculate $\psi_6 (r;R)$ for each particle where $r$ is the distance from the surface of an impurity particle.

Figure 8(a) shows  $\psi_6 (r;R)$ vs r data averaged over both the impurity particles for different value of R. We observe fluid like behavior of colloidal particles near the impurity walls, similar to the case of single impurity. There is an interfacial region of width $L= 6 $ before $\psi_6 (r;R)$ saturates. For larger $R$, $\psi_6 (r;R)$ saturates which we take as $Q_6$ for the system. Figure 8(b) shows the plot on the bond order parameter($Q_6)$ versus $R$. So far as $R\leq 2L$,  $Q_6$ decreases  with increasing $R$. $Q_6$ shows a discontinuity around R= 12. Beyond this value of $R$,  $Q_6$ increases abruptly to 0.70 after which there is not much change in $Q_6$. So far as $R<2L$, the liquid interface surrounding the individual particles overlap and destroy the order. 

Since the impurity doping is done on a fluid phase, our data suggest that a sufficiently large impurity induces QLRO phase when the impurity is doped in a bulk fluid phase.  Further, $Q_{6}<Q_{T}$ which is due to proliferation of defects, stabilized over the impurity surface, into the bulk. Smaller impurity has no effect on fluid structure. These observations are qualitatively consistent with observations on heterogeneous nucleation of colloidal crystal from fluid that nucleation is favoured by large impurities\cite{Lowen2015,sandomirski_review2014,Villeneuve2005jpcm,Villeneuve2005science}. In the presence of the second impurity, the quasi-long ranged order\cite{rjc,r59,r60,r63,r65} depends on the distance between the impurities. The mean distance among the quenched impurity should be larger than the twice interfacial width, L of the wetting layer about the impurity particle so that the order phase is not suppressed. At a finite impurity concentrations($\rho_i$), the impurity sites are distributed at mean separation(s)  $\sim \frac{1}{\sqrt\rho_i}$. The quasi-long ranged order is suppressed if $s \leq 2L$ . This corresponds to the threshold concentration $\rho_c =\frac{1}{4L^2}$. The QLRO will survive for concentration below this threshold. This is qualitatively consistent with the earlier experiments\cite{hutchinson2022prm,chen2021prl}, although these works consider anisotropic impurities. The impurity packing fraction threshold $\sim 0.17$ for our system. 

\begin{figure}[h]
\centering
    \includegraphics[height=4.0cm]{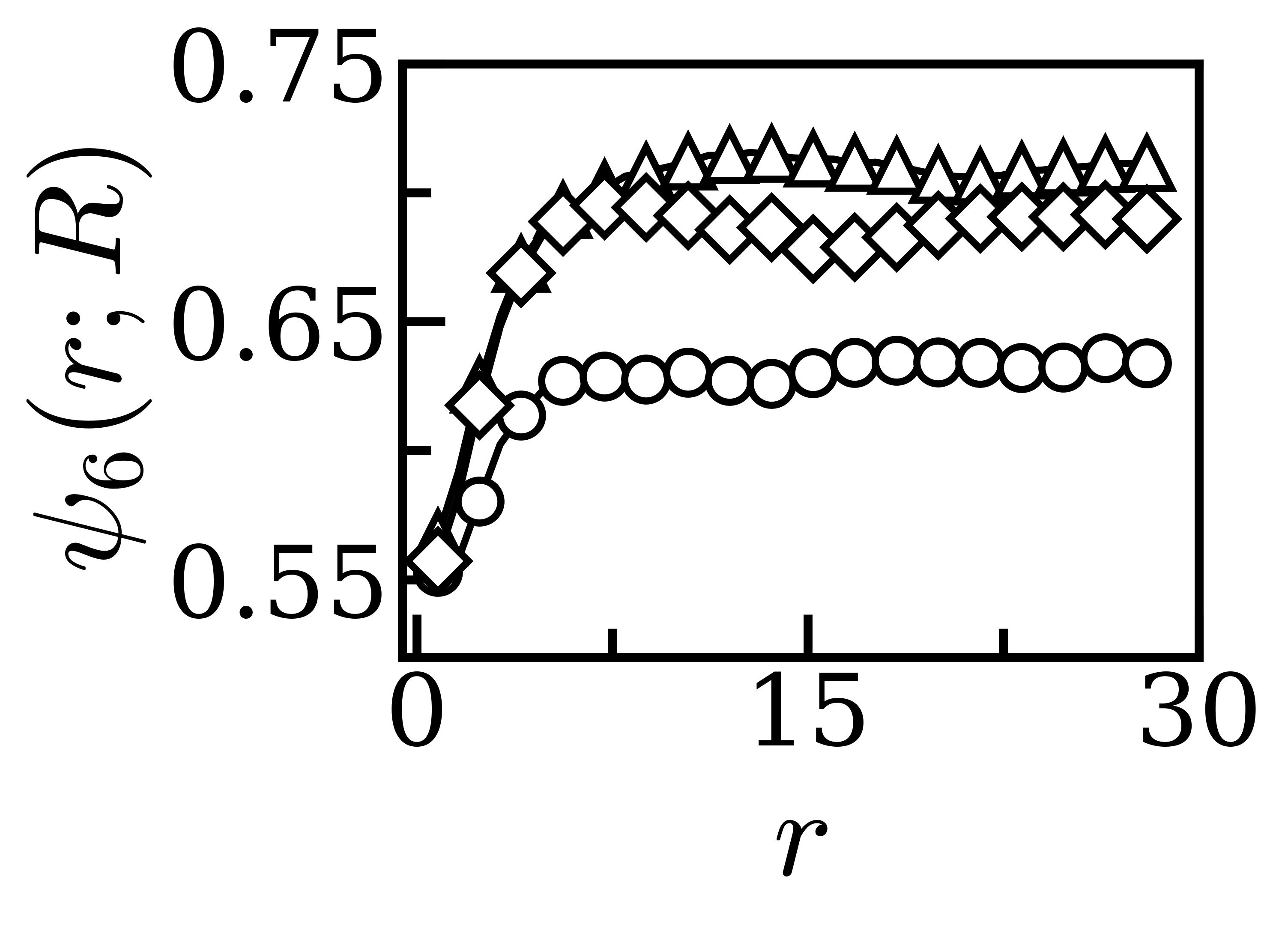}
        \put(-30,40){\textbf{(a)}}
    \includegraphics[height=3.9 cm]{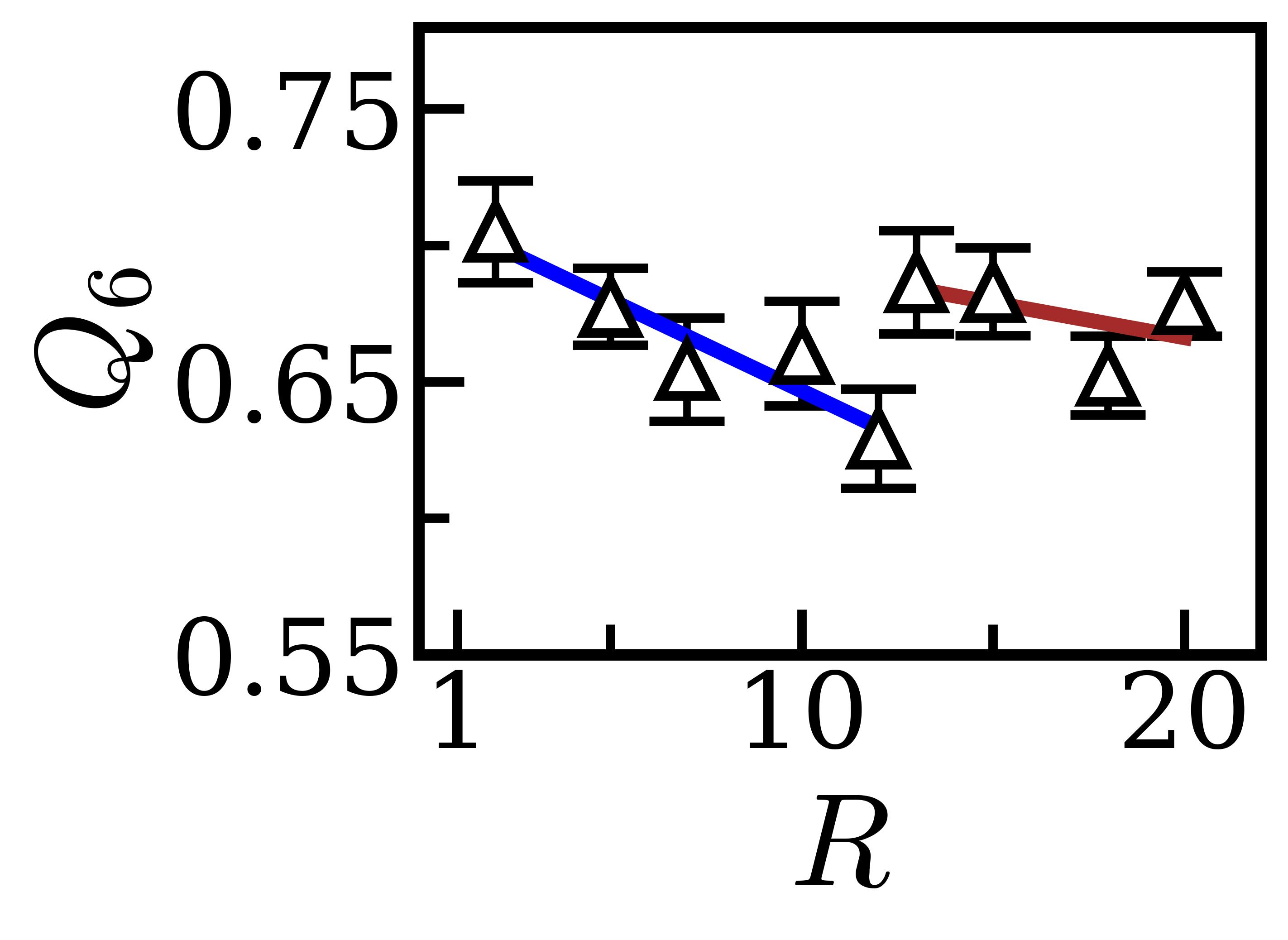}
        \put(-28,40){\textbf{(b)}}

    \caption{ \small{\textbf{(a)} $\psi_6(r,R)$ vs r for $\sigma_{imp} = 32$ for $R$= 2(triangle),R=12(circle), R=15(diamond). \textbf{(b)} $Q_6$ vs R plot. Solid lines are linear fit of the data.}}

\end{figure}
\section{Conclusion}

To summarize we carry out MD simulations to characterize the impurity-host fluid interface in colloidal fluid film in presence of a quenched impurity near the bulk fluid-crystal transition. Our simulations show that the surface of the impurity particles is always wet by a fluid layer, and the order in the system takes place over an interface of finite thickness for a sufficiently large impurity. The bulk state beyond the interface shows QLRO due to liquid-solid phase coexistence for sufficiently large impurity. The single particle dynamics is heterogeneous in the QLRO phase. Particles in the vicinity of impurity-host fluid interface are faster than particles deep in the bulk. The separation between the impurity sites must be larger than the interfacial width of the wetting layer over the impurity particles so as not to frustrate the ordered phase in the system in the presence of multiple impurities. Our results on the static and dynamic interfacial properties can be verified by direct video-microscopy on colloidal films.

\section*{Acknowledgments}
We sincerely acknowledge Prof. Punyabrata Pradhan and Dr. Suman Dutta for helpful discussions. JC acknowledges the CSIR for funding through CSIR- Emeritus scientist scheme.



\bibliography{rsc} 
\bibliographystyle{rsc} 

\end{document}